\documentclass[12pt]{article}
\usepackage{bbm}
\usepackage{color}
\usepackage{authblk}
\usepackage{latexsym}
\usepackage{epsfig,amssymb,euscript, mathrsfs,cite}
\usepackage{amsmath}
\usepackage{mathrsfs} 
\usepackage{enumerate}
\definecolor{MyDarkBlue}{rgb}{0.15,0.15,0.45}
\usepackage[linktocpage=true]{hyperref}
\hypersetup{
colorlinks=true,
citecolor=MyDarkBlue,
linkcolor=MyDarkBlue,
urlcolor=MyDarkBlue,
pdfauthor={},
pdftitle={},
pdfsubject={hep-th}
}
\newsavebox{\ns}
\newsavebox{\dbrane}
\newsavebox{\dbshort}

\def\be{\begin{equation}}
\def\ee{\end{equation}}
\def\bea{\begin{eqnarray}}
\def\eea{\end{eqnarray}}

\def\v{\vec{v}}

\newcommand{\nn}{\nonumber\\}

\newcommand\R{\mathbb{R}}
\newcommand\Z{\mathbb{Z}}

\newcommand\C{\mathbb{C}}

\newcommand\diff{\mathrm{d}}

\newcommand{\ii}{\mathrm{i}}

\newcommand{\ex}{\mathrm{e}}
\newcommand{\vol}{\mathrm{vol}}

\newcommand{\sss}{w}

\newcommand{\cZ}{\mathscr{Z}}
\newcommand{\csugra}{c_{\mathrm{sugra}}}

\newcommand{\Ssusy}{S_{\mathrm{SUSY}}}

\newcommand{\x}{x}

\newlength{\sswidth}

\newcommand{\pJ}{\mathtt{p}}
\newcommand{\qJ}{\mathtt{q}}
\newcommand{\X}{X}
\newcommand{\Bp}{k}

\newcommand{\n}{\mathtt{n}}
\newcommand{\m}{\mathtt{m}}

\numberwithin{equation}{section}       

\begin{document}

\begin{titlepage}

\begin{flushright}
Imperial/TP/2019/JG/03\\
\end{flushright}

\vskip 1.5cm

\begin{center}

{\Large \bf  Fibred GK geometry and }

\vskip 0.5cm

{\Large \bf supersymmetric $AdS$ solutions}

\vskip 1cm

{Jerome P. Gauntlett$^{\mathrm{a}}$, 
Dario Martelli$^{\mathrm{b,c,d}}$
and James Sparks$^{\mathrm{e}}$}

\vskip 0.5cm

${}^{\mathrm{a}}$\textit{Blackett Laboratory, Imperial College, \\
Prince Consort Rd., London, SW7 2AZ, U.K.\\}

\vskip 0.2cm

${}^{\mathrm{b}}$\textit{Dipartimento di Matematica ``Giuseppe Peano'', Universit\`a di Torino,\\
Via Carlo Alberto 10, 10123 Torino, Italy}

\vskip 0.2cm

${}^{\mathrm{c}}$\textit{INFN, Sezione di Torino \&}   ${}^{\mathrm{d}}$\textit{Arnold--Regge Center,\\
 Via Pietro Giuria 1, 10125 Torino, Italy}

\vskip 0.2 cm
${}^{\,\mathrm{e}}$\textit{Mathematical Institute, University of Oxford,\\
Andrew Wiles Building, Radcliffe Observatory Quarter,\\
Woodstock Road, Oxford, OX2 6GG, U.K.\\}

\end{center}

\vskip 1.5 cm

\begin{abstract}
\noindent  
We continue our study of a general class of $\mathcal{N}=2$ supersymmetric $AdS_3\times Y_7$ and $AdS_2\times Y_9$ solutions of type IIB and $D=11$ supergravity, respectively. The geometry of the internal spaces is part of a general family
of ``GK geometries", $Y_{2n+1}$, $n\ge 3$, and here we study examples in which $Y_{2n+1}$ fibres
over a K\"ahler base manifold $B_{2\Bp}$, with toric fibres. We show that the flux quantization conditions, and an action function that determines the supersymmetric $R$-symmetry Killing vector of a geometry, 
may all be written in terms of the  ``master volume'' of the fibre, together with certain global data associated with the K\"ahler base. In particular, 
this allows one to compute the central charge and entropy of the holographically dual $(0,2)$ SCFT and dual superconformal quantum mechanics, respectively, without 
knowing the explicit form of the $Y_7$ or $Y_9$ geometry. 
We illustrate with a number of examples, finding agreement with explicit supergravity solutions in cases where these are known, and we also obtain new results. In addition we present, {\it en passant}, new formulae for calculating the volumes of Sasaki-Einstein  manifolds.

\end{abstract}

\end{titlepage}

\pagestyle{plain}
\setcounter{page}{1}
\newcounter{bean}
\baselineskip17pt
\tableofcontents

\newpage


\section{Introduction}\label{sec:intro}

An interesting arena for exploring the AdS/CFT correspondence, both from the geometric and the field theory
points of view, is the class of supersymmetric $AdS_3\times Y_7$ solutions of type IIB supergravity of \cite{Kim:2005ez} and 
$AdS_2\times Y_9$ solutions of $D=11$ supergravity of \cite{Kim:2006qu}. These solutions are dual to $d=2$ SCFTs preserving $(0,2)$
supersymmetry and an
superconformal quantum mechanics preserving ${\cal N}=2$ supersymmetry, respectively, both of which have an abelian $R$-symmetry. The internal spaces of these supergravity solutions are low-dimensional examples of 
a novel kind of geometry, called ``GK geometry", which is defined on odd-dimensional manifolds $Y_{2n+1}$, $n\ge 3$ \cite{Gauntlett:2007ts}. GK geometry consists of a Riemannian metric, a scalar function $B$ and a closed two-form $F$ which 
extremizes a particular action and also admits a certain type of Killing spinor. Furthermore, motivated by the supergravity solutions, there is a natural flux quantization condition that can be imposed on cycles of co-dimension two. The GK geometries have a canonical $R$-symmetry Killing vector which, for the supergravity solutions, is precisely dual
to the $R$-symmetry in the dual field theory.

It has been shown recently  that the $R$-symmetry Killing vector in GK geometry can be obtained via an interesting variational
problem \cite{Couzens:2018wnk} that is analogous to the principle of volume minimization in Sasaki-Einstein geometry
\cite{Martelli:2005tp,Martelli:2006yb}. In the case of $n=3$, i.e. $Y_7$, this variational problem is a geometric realisation of
the $c$-extremization principle for the dual $(0,2)$ $d=2$ SCFTs proposed in \cite{Benini:2012cz} and allows one to obtain, for example, the central charge of the dual field theory without knowing the explicit $AdS_3\times Y_7$ solution. 
For the case of  $n=4$, i.e. $Y_9$, there is, in general, no analogous extremization principle in field theory
that one can compare with. However, for special subclasses of $Y_9$ it corresponds to the $I$-extremization principle of
\cite{Benini:2015eyy}, as shown in \cite{Gauntlett:2019roi,Hosseini:2019ddy}. 
Furthermore, when the $AdS_2\times Y_9$ solution arises as the near horizon limit of a black hole,
the geometric variational problem also allows one to calculate the entropy of the black hole, again without knowing the
explicit supergravity solution \cite{Couzens:2018wnk}. In particular, for the class of such black hole solutions that asymptotically
approach $AdS_4$, the connection with $I$-extremization provides a microscopic derivation of the black hole entropy, substantially extending \cite{Benini:2015eyy} (for other related work see, for example, \cite{Benini:2016rke,Cabo-Bizet:2017jsl,Azzurli:2017kxo,Benini:2017oxt,Hosseini:2017fjo}).

In previous work \cite{Gauntlett:2018dpc,Gauntlett:2019roi}, 
the variational problem of \cite{Couzens:2018wnk} was utilised to study specific classes of $Y_7$ and $Y_9$
that arise as a fibration over a Riemann surface $B_2=\Sigma_g$ with toric fibres $\X_5$ and $\X_7$, respectively.
By taking the $R$-symmetry Killing vector field to be tangent to the fibres it was shown that 
the general formulae in \cite{Couzens:2018wnk} can be recast in terms of a \emph{master volume} formula for the toric fibres 
which is a function of the toric data, a choice of $R$-symmetry vector and an arbitrary transverse K\"ahler class. 
It was shown that the extremization problem can be implemented using the master volume formula combined 
with a set of integers that determine the fibration of $\X_{5}$ or $\X_7$ over $\Sigma_g$, as well as a K\"ahler class parameter for $\Sigma_g$.

In this paper we substantially generalize these results. We will study the extremal problem for GK geometry on $Y_{2n+1}$
that arise as fibrations of the form $\X_{2r+1} \hookrightarrow  Y_{2r+2k+1}  \rightarrow  B_{2\Bp}$, with $r\ge 1$, $k\ge 1$ and $r+k=n\ge 3$.
We will assume that the base manifold $B_{2\Bp}$ of the fibration is K\"ahler, while the fibre is again taken to be toric. 
Remarkably, we will show that the extremal problem of \cite{Couzens:2018wnk} can again be implemented using the master volume formula for the toric fibres, as in the cases studied in \cite{Gauntlett:2018dpc,Gauntlett:2019roi}.  
One new feature is that while for $\Bp=1$ derivatives of the master volume with respect to the $R$-symmetry vector and 
K\"ahler class parameters appear, for $\Bp>1$ we will also need to consider derivatives with respect to the toric data.
We will present explicit formulae for specific values of $r,k$ that are associated with interesting $AdS_3$ and $AdS_2$ solutions,
but it is reasonably clear how to extend to other values. A simple explicit expression for the master volume formula in terms of the toric data
was given for $X_5$ and $X_7$ in \cite{Gauntlett:2018dpc,Gauntlett:2019roi}, respectively. Here we will also provide an analogous expression for the simpler case of $X_3$.

For application to the AdS/CFT correspondence the main utility of our new results is that one can calculate quantities of physical interest without knowing
the explicit supergravity solutions, just assuming that they exist. That being said, it is very satisfying to be able to
check the new formulae that we derive here against some explicitly known solutions. We will carry out such checks for
the class of $AdS_3\times Y_7$ solutions of type IIB found in \cite{Gauntlett:2006af}
with $\X_3\hookrightarrow Y_7\rightarrow B_4$, i.e. $r=1,\Bp=2$, with $B_4$ having a K\"ahler-Einstein metric. We will also carry out a similar check
for a class of $AdS_2\times Y_9$ solutions of $D=11$ supergravity with $\X_3\hookrightarrow Y_9\rightarrow B_6$,  i.e. $r=1,\Bp=3$, with $B_6$ having a K\"ahler-Einstein metric. These latter solutions were constructed in 
\cite{Gauntlett:2006ns} and here we complete the analysis of flux quantization.

The  plan  of  the  rest  of  the  paper  is  as  follows.  In section \ref{sec:GKsummary} we summarize general aspects of GK geometry and the associated extremal problem. In section \ref{sec:fibre} we discuss the toric 
fibres and their master volume. In section \ref{sec:fibredGK}, which contains our main new results, 
we present the formulae for implementing the extremal problem in the fibred GK geometries for K\"ahler base  manifolds of dimension $\Bp=1,2,3$.  We illustrate the formulae considering a variety of examples in section \ref{sec:examples}, focusing on the new cases of $\Bp=2$ and $\Bp=3$. In addition to reproducing the results of some known explicit supergravity solutions, where the bases $B_4$ and $B_6$ are K\"ahler-Einstein manifolds, we also work out examples where the base manifold is K\"ahler, but not Einstein. In particular, we consider $B_4=\Sigma_{g_1}\times \Sigma_{g_2}$, the product of two Riemann surfaces of genus $g_1$ and $g_2$, as well as $B_4=\mathbb{F}_n$, the $n$-th Hirzebruch surface. We conclude in section \ref{discsec} with some discussion.
The Appendices \ref{app:identities}--\ref{app:examples} contain the derivations of the various key identities involving the master volume, that we use in the main part of the paper. 
We have also included an Appendix \ref{app:Sas}, which explains how the formalism developed in \cite{Gauntlett:2019roi, Gauntlett:2018dpc} and the present paper allows one to efficiently compute the Sasakian volume function of \cite{Martelli:2005tp, Martelli:2006yb}.


\section{GK geometry and the extremal problem}\label{sec:GKsummary}

We begin by  briefly summarizing some aspects of GK geometry \cite{Gauntlett:2007ts}. This is a geometry defined on an odd-dimensional manifold, $Y_{2n+1}$, with $n\ge 3$, consisting of a metric, a scalar function $B$ and a closed two-form $F$, so that $\diff F=0$.

The existence of ``supersymmetry", by which we mean the
existence of certain Killing spinors given in \cite{Gauntlett:2007ts}, implies that the metric on $Y_{2n+1}$ has a unit norm Killing vector $\xi$, called the $R$-symmetry 
vector field. Since $\xi$ is nowhere vanishing it defines a foliation $\mathcal{F}_\xi$ of $Y_{2n+1}$. In local coordinates we may write 
\begin{align}
\xi = \frac{1}{c}\partial_z\, , \qquad \eta = c(\diff z + P)\, ,
\end{align}
where $c\equiv (n-2)/2$ and $\eta$ is the Killing one-form dual to $\xi$.
The metric on $Y_{2n+1}$ then has the form
\begin{align}\label{GKmetric}
\diff s^2_{2n+1}=\eta^2+\ex^B \diff s^2_{2n}\,,
\end{align}
where $\diff s^2_{2n}$ is a K\"ahler metric transverse to $\mathcal{F}_\xi$. This K\"ahler metric, with transverse K\"ahler two-form $J$,
Ricci two-form $\rho=\diff P$ and Ricci scalar $R$, determines all of the remaining fields. Specifically,
\begin{align}
\ex^B=\frac{c^2}{2}R\,,\qquad\qquad
F=-\frac{1}{c}J+\diff \left(\ex^{-B}\eta\right)\, .
\end{align}
In particular, notice that we require positive scalar curvature, $R>0$. 
These off-shell ``supersymmetric geometries" become on-shell GK geometries, or ``supersymmetric solutions", 
provided that the transverse K\"ahler metric satisfies the non-linear partial differential equation
\begin{align}\label{boxR}
\Box R &=\frac{1}{2}R^2 - R_{ij}R^{ij}~.
\end{align}

For $n=3$ and $n=4$ these give rise to supersymmetric $AdS_3\times Y_7$ and $AdS_2\times Y_9$ solutions of type IIB and $D=11$ supergravity,
which we describe in more detail below. For these cases we must impose a flux quantization condition for cycles of codimension two, 
and this naturally generalizes to all $n\geq 3$. Specifically, if $\Sigma_A$ are a basis for the free part of $H_{2n-1}(Y_{2n+1},\mathbb{Z})$
we impose
\begin{align}\label{fluxqc}
\int_{\Sigma_A}\left[\eta \wedge \rho\wedge \tfrac{1}{(n-2)!}J^{n-2}  + \frac{c}{2}*_{2n}\diff R\right]=\nu_{n} N_A\,,
\end{align}
where $N_A\in\mathbb{Z}$ and the non-zero, real constant $\nu_n$ is explicitly fixed only for the cases of $n=3$ and $n=4$, as given below.

We also recall that for an off-shell supersymmetric geometry (i.e. not imposing \eqref{boxR}) the real cone over 
$Y_{2n+1}$,  $C(Y_{2n+1})\equiv \R_{>0}\times Y_{2n+1}$, equipped with the conical metric
\begin{align}\label{metriccone}
\diff s^2_{2n+2} &= \diff \varrho^2 + \varrho^2 \diff s^2_{2n+1}~,
\end{align}
has some important properties. The cone $C(Y_{2n+1})$ has an integrable complex structure, and 
there exists a nowhere vanishing holomorphic $(n+1,0)$-form $\Psi$,
which, furthermore, is closed $\diff\Psi=0$. It follows that 
 $C(Y_{2n+1})$ has vanishing first Chern class. Additionally, the $R$-symmetry vector $\xi$ is holomorphic, 
and moreover $\Psi$ has a fixed charge with respect to the $R$-symmetry vector:
\begin{align}\label{chgepsi}
\mathcal{L}_\xi\Psi=\frac{\ii}{c}\Psi\,.
\end{align}

We can now summarize the extremal problem for the off-shell supersymmetric geometry that was presented in \cite{Couzens:2018wnk}.
We fix a complex cone $C(Y_{2n+1})=\R_{>0}\times Y_{2n+1}$ with holomorphic volume form $\Psi$, and holomorphic $U(1)^s$ action. 
We then choose a fiducial holomorphic $R$-symmetry vector $\xi$ and demand that the holomorphic volume form has fixed charge $1/c$, as in \eqref{chgepsi}. 
This choice of $\xi$ defines a foliation $\mathcal{F}_\xi$, and  we then further choose a transverse K\"ahler metric with 
basic cohomology class $[J]\in H^{1,1}_{{B}}(\mathcal{F}_\xi)$. We do not impose the condition \eqref{boxR}, as this would immediately put us 
on-shell. However, in order to impose the flux quantization condition (\ref{fluxqc}) we impose that the integral of \eqref{boxR} is satisfied. Specifically, we impose
the topological constraint
\begin{align}\label{constraint}
\ \int_{Y_{2n+1}}\eta\wedge \rho^2\wedge \tfrac{1}{(n-2)!}J^{n-2}=0\,,
 \end{align}
and also impose the flux quantization conditions\footnote{As discussed in \cite{Couzens:2018wnk}, this is equivalent to the flux quantization condition if
$H^2(Y_{2n+1},\R)   \cong  H^2_{{B}}(\mathcal{F}_\xi)/[\rho]$, which holds in the classes of examples studied in this paper.}
\begin{align}\label{quantize}
\int_{\Sigma_A}\eta\wedge \rho\wedge \tfrac{1}{(n-2)!}J^{n-2}&\ = \nu_n N_A\,,
\end{align}
with the basis of cycles $\{{\Sigma_A}\}~$ all tangent to $\xi$.
Finally,  an on-shell geometry, with properly quantized flux, extremizes the supersymmetric action
\begin{align}\label{susyact}
\Ssusy  \ &=  \ \int_{Y_{2n+1}}\eta\wedge \rho\wedge \tfrac{1}{(n-1)!}J^{n-1}\,.
 \end{align}
For a given $\xi$, it is important to emphasize that the quantities \eqref{constraint}, \eqref{quantize} and \eqref{susyact} just depend on the basic cohomology class $[J]\in H^{1,1}_{{B}}(\mathcal{F}_\xi)$, and not on $J$ itself \cite{Couzens:2018wnk}. Thus, for fixed $[J]$, we are extremizing over the space of $R$-symmetry vectors. A GK geometry with quantized flux 
is necessarily an extremal point, although as discussed in \cite{Couzens:2018wnk}, \cite{Gauntlett:2018dpc} for a given extremal point there may be obstructions to the existence of a 
corresponding GK supergravity solution, satisfying  (\ref{boxR}).

For the case of $n=3$, {i.e.} $Y_7$, the above extremal problem is associated to supersymmetric solutions of
type IIB supergravity of the form
\begin{align}
\diff s^2_{10} &= L^2 \ex^{-B/2}\left(\diff s^2({AdS_3}) + \diff s^2({Y_7})\right)~,\nn
F_5 &= -L^4\left(\vol_{{AdS}_3}\wedge F + *_7 F\right)~,\label{ansatz}
\end{align}
where $\diff s^2({{AdS}_3})$ has unit radius, and $L>0$ is a constant.
The five-form $F_5$ is properly quantized provided that we choose the constant $\nu_3$ to be
\begin{align}\label{nu3}
\nu_3=\frac{2(2\pi \ell_s)^4g_s}{L^4}\,, 
\end{align}
where $\ell_s$ is the string length, and $g_s$ is the constant string coupling.
Furthermore, the 
value of the on-shell action also determines the central charge, $\csugra$, of the dual field theory. 
Specifically, defining the ``trial central charge", $\cZ$, via
\begin{align}\label{cS}
\cZ  \equiv  \frac{3L^8}{(2\pi)^6g_s^2\ell_s^8} \Ssusy=\frac{12 (2\pi)^2}{\nu_3^2} \Ssusy~,
\end{align}
where $\Ssusy$ is the supersymmetric action (\ref{susyact}) with $n=3$, then we have
\begin{align}\label{cS2}
\cZ  |_\mathrm{on-shell}  =  \csugra~.
\end{align}

Similarly, when $n=4$, i.e. $Y_9$, the above extremal problem is associated to supersymmetric solutions of $D=11$ supergravity
of the form
\begin{align}\label{ansatzd11}
\diff s^2_{11} &= L^2 \ex^{-2B/3}\left(\diff s^2({{AdS}_2}) + \diff s^2(Y_{9})\right)~,\nn
G_4 &= L^3\vol_{{AdS}_2}\wedge F ~,
\end{align}
where $\diff s^2({{AdS}_2})$ has unit radius.
The four-form $G_4$ (or more precisely the Hodge dual seven-form $*_{11}G_4$) is properly quantized provided that we choose the constant $\nu_4$ to be
\begin{align}\label{nu4}
\nu_4=\frac{(2\pi \ell_p)^6}{L^6}\,,
\end{align}
where $\ell_p$ is the eleven-dimensional Planck length.
For this case we can define a ``trial entropy", $\mathscr{S}$, via
\begin{align}\label{cS2}
\mathscr{S}\equiv\frac{4\pi L^9}{(2\pi)^8\ell_p^9}\, \Ssusy~,
\end{align}
where $\Ssusy$ is the supersymmetric action (\ref{susyact}) with $n=4$.
In the case that the $D=11$ solution arises as the near-horizon limit of a supersymmetric black hole, it
is expected that $\mathscr{S}|_\mathrm{on-shell}$ is the entropy of the black hole \cite{Couzens:2018wnk}.
More generally, it is expected that $\mathscr{S}|_\mathrm{on-shell}$ is the logarithm of a supersymmetric partition function
of the dual quantum mechanical theory \cite{Couzens:2018wnk}. For the sub-class of solutions for which $Y_9$ is
the total space of a fibration of $\X_7$ over a Riemann surface there is also an established connection 
with $\mathcal{I}$-extremization \cite{Gauntlett:2019roi, Hosseini:2019ddy}, which provides
a state counting interpretation of the entropy of infinite classes of supersymmetric,
asymptotically $AdS_4$ black hole solutions.

In the remainder of the paper we will be interested in implementing the above extremal problem for 
geometries in which $Y_{2n+1}$ takes the fibred form
\begin{align}\label{fibred}
\X_{2r+1} \ \hookrightarrow \ Y_{2r+2\Bp+1} \ \rightarrow  B_{2\Bp}\, ,
\end{align}
where $n=r+\Bp$ and $B_{2\Bp}$ is a K\"ahler manifold. We will
further restrict to the case that the fibre manifold $\X_{2r+1}$ is toric, so that the cone metric over $\X_{2r+1}$ is invariant under 
a holomorphic $U(1)^{r+1}$ isometry, and moreover we take the $R$-symmetry vector $\xi$ to be tangent to the toric fibre. 
We describe this geometry in more detail in the next section.

\section{Geometry of the toric $\X_{2r+1}$ fibre}\label{sec:fibre}

In this section we describe the geometry of the fibres $\X_{2r+1}$ in (\ref{fibred}), in particular introducing the so-called
master volume $\mathcal{V}$. Our discussion here summarizes and generalizes section 3 of \cite{Gauntlett:2018dpc} from dimension $r=2$ to arbitrary dimension, 
and in addition we derive some new identities satisfied by the master volume that will be important  later in the paper.

\subsection{Toric K\"ahler cones}\label{sec:toric}

We start by assuming that we have a toric K\"ahler cone, ${C}(\X_{2r+1})$,
in real dimension $2(r+1)$. By definition
these are K\"ahler metrics of the conical form
\begin{align}\label{kcone}
\diff s^2_{C(\X_{2r+1})} &= \diff \varrho^2 + \varrho^2\diff s^2_{2r+1}~,
\end{align}
that admit a $U(1)^{r+1}$ action. This action is taken to be generated by the holomorphic Killing vectors
$\partial_{\varphi_i}$, $i=1,\dots, r+1$,  with each $\varphi_i$ having period $2\pi$. 
Moreover, we take $C(\X_{2r+1})$ to be 
Gorenstein, 
meaning that it admits a global 
holomorphic $(r+1,0)$-form $\Psi_{(r+1,0)}$. For convenience we choose a basis so that this holomorphic volume form has unit charge
under $\partial_{\varphi_1}$ and is uncharged under $\partial_{\varphi_i}$, $i=2,3,\ldots, r+1$. 

The manifold $\X_{2r+1}$ is embedded at $\varrho=1$. 
The complex structure of the cone pairs the radial vector $\varrho\partial_\varrho$ with the Killing vector field
$\xi$ tangent to $\X_{2r+1}$, which we may write as
\bea\label{Reebbasis}
\xi &=& \sum_{i=1}^{r+1} b_i\partial_{\varphi_i}~.
\eea 
The vector $\vec{b}=(b_1,\ldots,b_{r+1})$ then parametrizes the choice of $R$-symmetry vector $\xi$.
Notice that we then have 
\begin{align}\label{b1charge}
\mathcal{L}_\xi \Psi_{(r+1,0)} = \ii b_1 \Psi_{(r+1,0)}\, .
\end{align}

The complex structure likewise pairs the one-form $\eta$ dual to $\xi$ 
with $\diff \varrho/\varrho$. In particular for K\"ahler cones
\bea\label{detaSasakian}
\diff\eta &=& 2\omega_{\mathrm{Sasakian}}~,
\eea
where $\omega_{\mathrm{Sasakian}}$ is the transverse K\"ahler form. Because $\diff\eta$ 
is also a transverse symplectic form in this case, by definition $\eta$ is a 
contact
one-form on $\X_{2r+1}$. 
The unique vector field $\xi$ satisfying $\xi\lrcorner\eta=1$, $\xi\lrcorner \diff\eta=0$ is then 
also called the 
Reeb vector field. 
We may write the (Sasakian) metric on $\X_{2r+1}$ as
\bea\label{splitmetric}
\diff s^2_{2r+1} &=& \eta^2 + \diff s^2_{2r}({\omega})~,
\eea
where $\diff s^2_{2r}({\omega})$ is the transverse K\"ahler metric with K\"ahler form $\omega=\omega_{\mathrm{Sasakian}}$. 
We note that (\ref{b1charge}) implies that 
\begin{align}\label{bcoh}
[\diff\eta] = \frac{1}{b_1}[\rho]\in H^2_B(\mathcal{F}_\xi)\, , 
\end{align}
where $\mathcal{F}_\xi$ is the foliation of $\X_{2r+1}$ induced by the choice of Reeb vector $\xi$, and 
$\rho$ denotes the Ricci two-form of the transverse K\"ahler metric $\diff s^2_{2r}({\omega})$.

We may next  define the moment map coordinates
\bea\label{yi}
y_i & \equiv & \tfrac{1}{2}\varrho^2\partial_{\varphi_i}\lrcorner \eta~, \qquad i=1,\ldots,r+1~.
\eea
These span the so-called moment map polyhedral cone $\mathcal{C}\subset \R^{r+1}$, where the 
$\vec{y}=(y_1,\ldots,y_{r+1})$ are standard coordinates on $\R^{r+1}$. The polyhedral cone $\mathcal{C}$, which is convex, may be written as
\bea
\mathcal{C} &=& \{\vec{y}\in \R^{r+1}\ \mid \ (\vec{y},\v_a)\geq 0~, \quad a=1,\ldots, d\}~,
\eea
where $\v_a\in\Z^{r+1}$ are the inward pointing primitive normals to the facets of the polyhedral cone, and the index $a=1,\ldots,d\geq {r+1}$ labels the facets. 
Furthermore, $v_a=(1,w_a)$, where $w_a\in \Z^r$, follows from the Gorenstein condition in the basis for $U(1)^{r+1}$ described above. 

As shown in \cite{Martelli:2005tp}, for a K\"ahler cone metric on $C(\X_{2r+1})$ the $R$-symmetry vector $\vec{b}=(b_1,\ldots,b_{r+1})$ lies 
in the interior of the Reeb cone, $\vec{b}\in \mathcal{C}_{\mathrm{int}}^*$. Here the 
Reeb cone
$\mathcal{C}^*$ is defined to be the dual cone 
to $\mathcal{C}$, with $\mathcal{C}^*_{\mathrm{int}}$ being its open interior. Using $\xi\lrcorner \eta=1$, together with (\ref{Reebbasis}) and 
(\ref{yi}), the image of $\X_{2r+1}=\{\varrho=1\}$ under the moment map is then the compact, convex $r$-dimensional polytope
\begin{align}
P = P(\vec{b}) = \mathcal{C}\cap H(\vec{b})\, ,
\end{align}
where the 
Reeb hyperplane
is by definition
\begin{align}\label{ReebH}
H=H(\vec{b}) \equiv \left\{\vec{y}\in \R^{r+1}\mid (\vec{y},\vec{b}) = \tfrac{1}{2}\right\}\, .
\end{align}

\subsection{The master volume}\label{sec:master}

Following \cite{Gauntlett:2018dpc}, we first fix a choice of toric K\"ahler cone metric on the complex cone $C(\X_{2r+1})$. As described in the previous subsection, this allows us to introduce 
the moment map coordinates $\vec{y}$ in (\ref{yi}), together with the angular coordinates $\varphi_i$, $i=1,\ldots,r+1$, as coordinates on $C(\X_{2r+1})$. 
Geometrically, $C(\X_{2r+1})$ then fibres over the polyhedral cone $\mathcal{C}$: over the interior $\mathcal{C}_{\mathrm{int}}$ 
of  $\mathcal{C}$ this is a trivial $U(1)^{r+1}$ fibration, with the normal vectors $\v_a\in \Z^{r+1}$ to each 
bounding facet $\{(\vec{y},\vec{v}_a)=0\}\subset \partial\mathcal{C}$ specifying which $U(1)\subset U(1)^{r+1}$ collapses along that facet.  
Each such facet is also the image under the moment map of a 
toric divisor
in $C(\X_{2r+1})$ -- that is, a complex codimension one submanifold that is invariant under 
the torus $U(1)^{r+1}$. The index $a=1,\ldots,d$ thus also labels the toric divisors.

For a fixed choice 
of such complex cone, with Reeb vector $\xi$ given by (\ref{Reebbasis}),
we would then like to study a more general class of transversely K\"ahler metrics of the form (\ref{splitmetric}). In particular, we are interested in the ``master volume'' defined by
\begin{align}
\label{defVmaaster}
\mathcal{V} \equiv   \int_{\X_{2r+1}}\eta\wedge \tfrac{1}{r!}\omega^r~,
\end{align}
as a function both of the vector $\xi$, and transverse K\"ahler class $[\omega]\in H^2_B(\mathcal{F}_\xi)$.
Following \cite{Gauntlett:2018dpc}, if we take
$c_a$ to be basic representatives in $H^2_B(\mathcal{F}_\xi)$ that lift to 
 integral classes in $H^2(\X_{2r+1},\mathbb{Z})$, which are Poincar\'e dual to 
the restriction of the toric divisors on $C(\X_{2r+1})$, then we can write
\begin{align}\label{omegalambda}
[\omega] = -2\pi\sum_{a=1}^d \lambda_a c_a \in  H^2_B(\mathcal{F}_\xi) \, . 
\end{align}
The $c_a$ are not all independent and $[\omega]$ in fact only depends on $d-r$  of the  $d$  parameters $\{ \lambda_a\}$, as 
we shall see shortly. 
It will also be useful to note that the first Chern class of the foliation can be written in terms of the $c_a$ as 
\begin{align}\label{rhoca}
[\rho] &= 2\pi\sum_{a=1}^d c_a  \in H^2_B(\mathcal{F}_\xi)~.
\end{align}
In the special case in which 
\be
\label{lambdaSas}
\lambda_a  =  -\frac{1}{2b_1}\,, \qquad a=1,\dots d\,,
\ee
we recover the Sasakian  K\"ahler class $[\rho]= 2 b_1 [\omega_\mathrm{Sasakian}]$ and the master volume (\ref{defVmaaster})
reduces to the Sasakian volume.

Again 
following \cite{Gauntlett:2018dpc}, the master volume (\ref{defVmaaster}) may be written as
\begin{align}\label{VEuc}
\mathcal{V} = \frac{(2\pi)^{r+1}}{|\vec{b}|}\vol(\mathcal{P})~.
\end{align}
Here the factor of $(2\pi)^{r+1}$ arises by integrating over the torus $U(1)^{r+1}$, while $\vol (\mathcal{P})$ is 
the Euclidean volume of the compact, convex $r$-dimensional polytope
\begin{align}\label{generalP}
\mathcal{P} = \mathcal{P}(\vec{b};\{\lambda_a\}) \ \equiv \ \{\vec{y}\in H(\vec{b}) \ \mid \ (\vec{y}-\vec{y}^{(0)},\v_a) \geq \lambda_a~, \quad a=1,\ldots,d\}~.
\end{align}
Here
\begin{align}\label{originP}
\vec{y}^{(0)} \equiv \left(\frac{1}{2b_1},0,\ldots ,0\right) \in H~,
\end{align}
which lies in the interior of $\mathcal{P}$, 
while the $\{\lambda_a\}$ parameters determine the transverse K\"ahler class. We next introduce 
the new coordinates
\begin{align}\label{xi}
x_i \equiv y_i - y^{(0)}_i\, .
\end{align}
Notice that the inequalities defining the polytope $\mathcal{P}$ then become simply $(\vec{x},\vec{v}_a)\geq \lambda_a$, $a=1,\ldots, d$, which 
is the usual way the moment polytope is presented in toric K\"ahler geometry. In this case $x_i$ is the Hamiltonian function
for the $i$th $U(1)$ Killing vector $\partial_{\varphi_i}$ with respect to the (transverse) K\"ahler form $\omega$, i.e. $\diff x_i=-\partial_{\varphi_i}\lrcorner\omega$.
Using \eqref{xi}
we may then also write the master volume (\ref{VEuc}) as an integral
\begin{align}\label{Vthetadelta}
\mathcal{V} = \mathcal{V}(\vec{b};\{\lambda_a\};\{\vec{v}_a\}) = (2\pi)^{r+1}\int_{\R^{r+1}} \prod_{a=1}^d \theta((\vec{\x},\v_a)-\lambda_a) \delta((\vec{\x},\vec{b}))\, ,
\end{align} 
where the integration over $\R^{r+1}$ uses the standard Euclidean measure $\diff x_1\wedge\cdots\wedge \diff x_{r+1}$.
Here we have emphasized in the notation that the master volume also depends on the choice of polyhedral cone $\mathcal{C}$, via its primitive normal vectors $\vec{v}_a\in\Z^{r+1}$, 
as well as the choice of $R$-symmetry vector $\vec{b}$ and K\"ahler class parameters $\{\lambda_a\}$. 
Using (\ref{Vthetadelta}) it is shown in Appendix \ref{app:identities} that $\mathcal{V}$ satisfies the identity
\begin{align}\label{keyvRrel}
\sum_{a=1}^d\left(\vec{v}_a-\frac{\vec{b}}{b_1}\right) \frac{\partial \mathcal{V}}{\partial \lambda_a}= 0\,,
\end{align}
meaning that this equation holds for all $\vec{b}$ and $\{\lambda_a\}$ (for fixed polyhedral cone and hence fixed $\{\vec{v}_a\}$). It follows that the
master volume is invariant under the ``gauge" transformations 
\begin{align}\label{lamgt}
\lambda_a\ \to\ \lambda_a+\sum_{i=1}^{r+1}\gamma_i(v_a^i b_1-b_i)\,,
\end{align}
for arbitrary constants $\gamma_i$, generalizing a result of \cite{Hosseini:2019use}. 
For $\X_{2r+1}$, noting that the transformation parametrized by $\gamma_1$ is trivial, this explicitly shows that
the master volume only depends on $d-r$ of the $d$ parameters $\{\lambda_a\}$.

The master volume $ \mathcal{V} $ is  homogeneous of degree $r$ in the $\lambda_a$, and
we have
 \begin{align}\label{mvapphere}
\mathcal{V} \ \equiv  \ \int_{\X_{2r+1}}\eta\wedge \tfrac{1}{r!}  \omega^r= (-2\pi)^r\sum_{a_1,\dots, a_r=1}^d \tfrac{1}{r!}I_{a_1\dots a_r} \lambda_{a_1}\dots \lambda_{a_r}~,
 \end{align}
 where the ``intersection numbers'' $I_{a_1\dots a_r}$ are defined as
\begin{align}\label{intersection}
 I_{a_1\dots a_r} \ \equiv \  \int_{Y_{2r+1}}\eta\wedge c_{a_1}\wedge\dots  \wedge c_{a_r} &=  \frac{1}{(-2\pi)^r}\frac{\partial^r\mathcal{V}}{\partial\lambda_{a_1}\dots  \partial\lambda_{a_r}}~.
\end{align}
We may then calculate
\begin{align}\label{lamderv}
 \int_{\X_{2r+1}}\eta\wedge \rho^s \wedge \tfrac{1}{(r-s)!}{\omega^{r-s}}
 =(-1)^s\sum_{a_1,\dots, a_s=1}^d \frac{\partial^s\mathcal{V}}{\partial\lambda_{a_1}\dots \partial\lambda_{a_s}}
\,.
 \end{align}
We also are interested in integrating over $S_a$, 
the $(2r-1)$-cycle in $\X_{2r+1}$ associated with a toric divisor on the cone and Poincar\'e dual to $c_a$.
We have
\begin{align}\label{SaVnew}
\int_{S_a} \eta\wedge \rho^s\wedge\tfrac{1}{(r-s-1)!}\omega^{r-s-1} &= \int_{Y_{2r+1}}\eta\wedge \rho^s\wedge\tfrac{1}{(r-s-1)!}\omega^{r-s-1} \wedge c_a \nn
&=\frac{(-1)^{s+1}}{2\pi}\sum_{b_1,\dots,b_s=1}^d \frac{\partial^{s+1} \mathcal{V}}{\partial \lambda_a\partial \lambda_{b_1}\dots\partial\lambda_{b_s}}\,.
\end{align}
In Appendix \ref{app:identities} we also show that the
master volume $ \mathcal{V} $ is homogeneous of degree $-1$ in the $b_i$.

It is possible to obtain very explicit formulas for the master volume in low dimensions. In dimensions $r=2$ and $r=3$ the relevant formulae for $\X_5$ and $\X_7$ were 
derived in \cite{Gauntlett:2018dpc} and \cite{Gauntlett:2019roi}, respectively. In the present paper we shall also be interested in the case $r=1$, with a three-dimensional toric fibre $\X_3$. 
In this case the toric data of a Gorenstein K\"ahler cone of complex dimension $r+1=2$ is given by the
two inward pointing normal vectors $v_1=(1,0)$, $v_2=(1,p)$, where $p\in\mathbb{N}$. 
 This describes an $A_{p-1}$ singularity, $C(\X_3)=\C^2/\Z_p$, with the $\Z_p$ 
action on $\C^2$ given by $(z_1,z_2)\mapsto (\omega_p z_1,\omega_p^{-1}z_2)$, where $\omega_p$ is a primitive $p$th root of unity. 
As shown in Appendix \ref{app:X3}, the master volume of $\X_3$ in this case is simply
\begin{align}\label{VX3}
\mathcal{V}(\vec{b};\lambda_1,\lambda_2;\vec{v}_1,\vec{v}_2) =  (2\pi)^2\sum_{a=1}^2(-1)^a \frac{\lambda_a}{[\vec{v}_a,\vec{b}]}~,
\end{align}
where here $[\vec{v}_a,\vec{b}]$ denotes the determinant of the $2\times 2$ matrix, i.e. $[\vec{v}_a,\vec{b}]\equiv  \varepsilon_{ij}v_a^i b^j$.
Later in the paper we will also need the master volume in dimension $r=2$. In this case the master volume of $\X_5$ is \cite{Gauntlett:2018dpc}
\begin{align}
\mathcal{V}(\vec{b};\{\lambda_a\};\{\vec{v}_a\})  =  \frac{(2\pi)^3}{2}\sum_{a=1}^d \lambda_a \frac{\lambda_{a-1}[\v_a,\v_{a+1},\vec{b}] - \lambda_a [\v_{a-1},\v_{a+1},\vec{b}]+\lambda_{a+1}[\v_{a-1},\v_a,\vec{b}]}{[\v_{a-1},\v_a,\vec{b}][\v_a,\v_{a+1},\vec{b}]}~,\nonumber 
\end{align}
where $[\cdot,\cdot,\cdot]$  denotes a $3\times 3$ determinant. Here the facets are ordered anti-clockwise around the polyhedral cone, 
and we cyclically identify $\v_{d+1}\equiv \v_1$, $\v_0\equiv \v_d$, and similarly $\lambda_{d+1}\equiv \lambda_1$, $\lambda_0\equiv \lambda_d$. 

Finally, we note that the  formulae in this section assume that the polyhedral
cone $\mathcal{C}$ is convex, since we started the section with a cone that 
 admits a toric K\"ahler
cone metric. However, as discussed in \cite{Couzens:2018wnk, Gauntlett:2018dpc}, this convexity
condition is, in general, too restrictive for applications to the classes of $AdS_2$ and $AdS_3$
solutions of interest. Indeed, many such explicit supergravity solutions are associated
with ``non-convex toric cones'', as defined in \cite{Couzens:2018wnk}, which in particular have toric data
which do not define a convex polyhedral cone. As in the above papers and \cite{Gauntlett:2019roi}, we conjecture that the 
 key formulae in
this section are also applicable to non-convex toric cones, and we will assume that this
is the case in the sequel. The consistent picture that emerges, combined with similar
results in \cite{Couzens:2018wnk, Gauntlett:2018dpc, Gauntlett:2019roi}, strongly supports the validity of this conjecture.


\section{Fibred GK geometry}\label{sec:fibredGK}

We would like to study GK geometries of the fibred form (\ref{fibred}), where the fibres $\X_{2r+1}$ take the toric 
form described in section \ref{sec:fibre}. In particular, we would like to evaluate the constraint, flux quantization condition and 
supersymmetric action (\ref{constraint}), (\ref{quantize}), (\ref{susyact}) for these fibred geometries, respectively. In this section 
we follow a similar analysis to that in section 4 of \cite{Gauntlett:2018dpc}, which studied the case of 
$\X_5$ fibred over a Riemann surface $\Sigma_g$ of genus $g$. Extending this to $\X_{2r+1}$ fibred over 
a K\"ahler base $B_{2\Bp}$ is relatively straightforward, although for $\Bp>1$ 
various new features arise compared to the Riemann surface $\Bp=1$ case.

\subsection{General set-up}\label{sec:setup}

The manifolds $\X_{2r+1}$ by definition admit an isometric $U(1)^{r+1}$ action. We may use this symmetry to fibre $\X_{2r+1}$ over 
the K\"ahler base $B_{2\Bp}$ by picking $r+1$ $U(1)$ gauge fields $A_i$  on $B_{2\Bp}$, $i=1,\ldots,r+1$, with curvatures $F_i=\diff A_i$ 
given by
\begin{align}\label{Fi}
\frac{F_i}{2\pi}=\sum_\alpha n_{i}^{\alpha}c^{(2)}_\alpha\, .
\end{align}
Here $c^{(2)}_\alpha\in H^2(B_{2\Bp},\R)$ are closed two-forms that generate the free part of 
$H^2(B_{2\Bp},\Z)$, and $n_i^\alpha\in \Z$.  It will be convenient later to take 
the $c^{(2)}_\alpha\in H^2(B_{2\Bp},\R)$ to be Poincar\'e duals to a corresponding 
basis of $(2p-2)$-cycles $C^{(2p-2)}_\alpha\in H_{2p-2}(B_{2\Bp},\Z)$, which by definition means that
\begin{align}
\int_{C^{(2p-2)}_\alpha} \Phi = \int_{B_{2\Bp}} \Phi \wedge c^{(2)}_\alpha
\end{align}
holds for all closed $(2p-2)$-forms $\Phi$ on $B_{2\Bp}$. Having chosen the curvatures in (\ref{Fi}), which amounts to a choice 
of the integers $n_i^\alpha$, one then uses the corresponding $U(1)^{r+1}$ transition functions to fibre 
$\X_{2r+1}$ over $B_{2\Bp}$, using the toric action of $U(1)^{r+1}$  on $\X_{2r+1}$.\footnote{Notice 
that a choice of $n_i^\alpha\in \Z$ only determines the principal $U(1)^{r+1}$ bundle up to a torsion class in 
$H^2(B_{2\Bp},\Z)$, although this torsion data will not enter the formulae that follow.} 

More concretely, the above fibration amounts to a replacement
\begin{align}
\diff\varphi_i \rightarrow \diff\varphi_i + A_i\, \qquad i=1,\ldots,r+1\, ,
\end{align}
where recall that $\varphi_i$ are the $(2\pi)$-periodic coordinates on the torus $U(1)^{r+1}$. As in \cite{Gauntlett:2018dpc}, 
it is important here to emphasize that the quantities (\ref{constraint}), (\ref{quantize}), (\ref{susyact}) of interest 
on the total space of the fibration depend only on basic cohomology classes in $H^2_B(\mathcal{F}_\xi)$. This means that we may use any convenient representative of 
the various differential forms that enter these quantities -- we must only ensure that the representative we use has the correct 
basic cohomology class. 

With these comments in mind, after the fibration the contact one-form $\eta$ on the fibres $\X_{2r+1}$ is effectively replaced by 
\begin{align}\label{etatwisted}
\eta \rightarrow  \eta_{\mathrm{twisted}} & \equiv  2\sum_{i=1}^{r+1} \sss_i (\diff\varphi_i + A_i)~,
\end{align}
where we have defined
\begin{align}
w_i \equiv \left. y_i\right|_{\varrho=1} = \tfrac{1}{2}\partial_{\varphi_i}\lrcorner\eta\, .
\end{align}
Recall here that the $y_i$ are the moment map coordinates (\ref{yi}) on $C(\X_{2r+1})$, and $\X_{2r+1}=\{\varrho=1\}\subset C(\X_{2r+1})$.
We then have
\begin{align}
\diff\eta_{\mathrm{twisted}} &= 2\sum_{i=1}^{r+1} \diff \sss_i \wedge (\diff\varphi_i + A_i) + 2\sum_{i=1}^{r+1} \sss_i F_i\, .
\end{align}
For the transverse K\"ahler form $J$ we may write
\begin{align}\label{Jtwisted}
J &= \omega_{\mathrm{twisted}} + J_{B_{2\Bp}} + \mbox{basic exact}~,
\end{align}
up to an irrelevant basic exact form, 
where $J_{B_{2\Bp}}$ is a K\"ahler form on the base
$B_{2\Bp}$ and
\begin{align}\label{omegatwisted}
\omega_{\mathrm{twisted}}  \equiv & \sum_{i=1}^{r+1} \diff x_i \wedge (\diff\varphi_i + A_i) + \sum_{i=1}^{r+1} x_i F_i~.
\end{align}
Here we have identified 
\begin{align}\label{Hamilton}
\diff x_i = -\partial_{\varphi_i}\lrcorner \omega\, ,
\end{align}
so that the $x_i$ are global Hamiltonian functions on the fibre $\X_{2r+1}$, invariant under the torus action, {\it cf}. the discussion 
 after equation \eqref{xi}, where the same functions $x_i$ appear. 
Notice that these are a priori defined only up to an additive constant, but that via equation \eqref{Jtwisted} such a constant shift may be absorbed into a redefinition of
the K\"ahler form $J_{B_{2\Bp}}$. 
As in (\ref{Fi}) we may then similarly 
decompose this K\"ahler form on the base as
\begin{align}\label{aalphadef}
J_{B_{2\Bp}} = \sum_\alpha a^\alpha c^{(2)}_\alpha\, ,
\end{align}
where $a_\alpha\in \mathbb{R}$. 

At the level of the formulae  (\ref{constraint}), (\ref{quantize}), (\ref{susyact}), which are expressed as integrals on the total space of the fibration, 
we may then simply substitute
\begin{align}\label{subst}
J&\to \omega+x_iF_i+J_{B_{2\Bp}}\,,\nn
\eta&\to\eta\,,\nn
\diff\eta&\to \diff\eta+2w_iF_i\,,
\end{align}
where on the right hand side, in a slight abuse of notation, $\omega, \eta$ and $\diff\eta$ are quantities on the fibre $\X_{2r+1}$, 
while the remaining terms are quantities on the base $B_{2\Bp}$. 

The holomorphic $(r+1,0)$-form $\Psi_{(r+1,0)}$ on the cone $C(\X_{2r+1})$ over the fibre has unit charge under $\partial_{\varphi_1}$, 
meaning there is an explicit $\ex^{\ii\varphi_1}$ dependence,  
where recall that we have chosen the basis for the torus action so that this is the case. On the other hand, the holomorphic $(n+1,0)$-form 
$\Psi$ on $C(Y_{2n+1})$ is constructed by taking the wedge product of the canonical holomorphic $(\Bp,0)$-form on the K\"ahler base 
$B_{2\Bp}$ with the $(r+1,0)$-form $\Psi_{(r+1,0)}$ on the fibre, twisting the latter using the $r+1$ line bundles over $\Bp$ 
with curvatures $F_i$, $i=1,\ldots,r+1$. The canonical $(\Bp,0)$-form on the K\"ahler base $B_{2\Bp}$ is not globally defined in general (unless the 
base is Calabi-Yau), being a section of the canonical line bundle $K_{B_{2\Bp}}$. However, due to the twisting, $\ex^{\ii \varphi_1}$ 
is precisely a section of the line bundle over $B_{2\Bp}$ with first Chern class $[F_1/2\pi]\in H^2(B_{2\Bp},\Z)$.  Neither section exists globally 
in general, but the wedge product does have a global nowhere zero section, and hence gives rise to a global $(n+1,0)$-form $\Psi$ on 
$C(Y_{2n+1})$, precisely if
\begin{align}\label{twist1}
\left[\frac{F_1}{2\pi}\right] = -c_1(K_{B_{2\Bp}}) = c_1(B_{2\Bp})\, .
\end{align}
When this condition holds, the cone $C(Y_{2n+1})$ has a global $(n+1,0)$-form. 
The condition (\ref{twist1}) generalizes the twist condition over a Riemann surface (where $\Bp=1$) 
presented in \cite{Gauntlett:2018dpc, Gauntlett:2019roi}.

Finally, recalling \eqref{b1charge} and \eqref{chgepsi}, for $Y_{2r+2k+1}$ we need to take 
\begin{align}\label{b1form}
b_1=\frac{2}{r+k-2}\,.
\end{align}
In the expressions given in the next subsections, it is important that this condition is only imposed after taking any derivatives with respect to the $b_i$.

In the remainder of this section we simply present the final formulae for various low-dimensional cases of interest, referring to the appendices 
for further details of the calculations involved. 

\subsection{$\X_{2r+1} \ \hookrightarrow \ Y_{2r+3} \ \rightarrow  B_2$}\label{beetwo}

Here $r=2$ and $r=3$ are relevant for the type IIB case and the $D=11$ case, respectively.
In fact these two cases were already treated in 
\cite{Gauntlett:2018dpc} and \cite{Gauntlett:2019roi}, respectively. Generalizing the calculations to general
$r\geq 2$ is straightforward.

We begin by noting that the topological constraint \eqref{constraint} can be written in the form
\begin{align}\label{constraintagain}
 \sum_{a,b=1}^d \frac{\partial^2\mathcal{V}}{\partial\lambda_a\partial\lambda_b}\vol(B_2) 
   + b_1 \sum_{a=1}^d \sum_{i=1}^{r+1} \frac{\partial^2 \mathcal{V}}{\partial \lambda_a\partial b_i}\int_{B_2}F_i
    -  \sum_{a=1}^d\frac{\partial \mathcal{V}}{\partial\lambda_a}\int_{B_2}F_1=0~.
\end{align}
Next we consider the flux quantization conditions given in \eqref{quantize}. There are two classes of $(2r+1)$-cycles to consider.
First, there is the distinguished $(2r+1)$-cycle, $\Sigma$, obtained by picking a point on the K\"ahler base $B_2$. We find
\begin{align}\label{Nnice}
 -\sum_{a=1}^d\frac{\partial\mathcal{V}}{\partial\lambda_a}=\nu_{r+1}N~,
\end{align}
where we recall that $\nu_{r+1}$ is a non-zero, real constant, fixed for the case of $r=2,3$ as in \eqref{nu3}, \eqref{nu4}, respectively, and $N\in\mathbb{Z}$.
The second class of $(2r+1)$-cycles are given by the total spaces $\Sigma_a$ of the fibrations
\begin{align}
S_a \ \hookrightarrow \ \Sigma_a \ \rightarrow  B_2\, ,
\end{align}
where $S_a$ is a $(2r-1)$-cycle in $\X_{2r+1}$ associated with a toric divisor on the associated cone $C(\X_{2r+1})$. For these cycles we have
\begin{align}\label{Maint}
\frac{1}{2\pi} \sum_{b=1}^d \frac{\partial^2\mathcal{V}}{\partial\lambda_a\partial\lambda_b}\vol(B_2) 
+ \frac{b_1}{2\pi} \sum_{i=1}^{r+1}  \frac{\partial^2 \mathcal{V}}{\partial\lambda_a\partial b_i}\int_{B_2}F_i=\nu_{r+1}M_a ~,
\end{align}
with $M_a\in\mathbb{Z}$. The $S_a$ are not linearly independent cycles in the fibre $\X_{2r+1}$, which leads to the corresponding linear relations
among the flux numbers \cite{Gauntlett:2018dpc, Gauntlett:2019roi}:
\begin{align}
\sum_{a=1}^d v_a^i M_a = -N \int_{B_2} \frac{F_i}{2\pi}\, , \qquad i=1,\ldots,r+1\, .
\end{align}
In the above expressions, from \eqref{b1form}, we should take
\begin{align}
b_1=\frac{2}{r-1}\,,
\end{align}
after taking derivatives with respect to $b_i$.
Finally, the supersymmetric action, given in \eqref{susyact}, can be cast in the form
\begin{align}\label{susactnice1}
\Ssusy  =
\nu_{r+1}\frac{2\pi }{r}\left(
\frac{N}{2\pi} \vol(B_2) -\sum_{a=1}^d\lambda_a{M_a}
\right)\,.
\end{align}

\subsection{$\X_{2r+1} \ \hookrightarrow \ Y_{2r+5} \ \rightarrow  B_4$}\label{beefour}

Now $r=1$ is relevant for the type IIB case, while $r=2$ is relevant for the $D=11$ case.

The topological constraint condition \eqref{constraint} is given by
\begin{align}\label{constraintB4}
&\sum_{a,b=1}^d  \frac{\partial^2\mathcal{V}}{\partial\lambda_a\partial\lambda_b} \vol(B_4)
+b_1\sum_{i=1}^{r+1}\sum_{a=1}^d \frac{\partial^2\mathcal{V}}{\partial\lambda_a\partial b_i}\int_{B_4}F_i\wedge J_{B_{4}}
-\sum_{a=1}^d \frac{\partial\mathcal{V}}{\partial\lambda_a}\int_{B_4}F_1\wedge J_{B_{4}}\nn
&\qquad+{b_1^2}\sum_{i,j=1}^{r+1}\frac{\partial^2 \mathcal{V}}{\partial b_i\partial b_j}\int_{B_4}\tfrac{1}{2}F_i\wedge F_j = 0\,.
\end{align}
There are two types of flux integrals, corresponding to two types of $(2r+3)$-cycles. 
The first type of cycles have the fibred form 
\begin{align}
\X_{2r+1} \ \hookrightarrow \ \Sigma_\alpha \ \rightarrow C^{(2)}_\alpha\, ,
\end{align}
 with 
$C^{(2)}_\alpha\subset B_4$ a two-cycle. We find
\begin{align}\label{NalphaB4}
-\sum_{a=1}^d  \frac{\partial  \mathcal{V}}{\partial \lambda_a } \int_{C^{(2)}_\alpha}J_{B_{4}}
-b_1\sum_{i=1}^{r+1}\frac{\partial  \mathcal{V}}{\partial b_i }\int_{C^{(2)}_\alpha}F_i
=\nu_{r+2} N_\alpha
\,,
\end{align}
where $\nu_{r+2}$ is a non-zero, real constant, fixed for the case of $r=1,2$ as in \eqref{nu3}, \eqref{nu4}, respectively, and $N_\alpha\in\mathbb{Z}$. The second set of $(2r+3)$-cycles have the fibred form
\begin{align}
S_a \ \hookrightarrow \ \Sigma_a \ \rightarrow  B_4\, ,
\end{align}
where $S_a$ is a $(2r-1)$-cycle in $\X_{2r+1}$ associated with a toric divisor on the associated cone $C(\X_{2r+1})$.
We find
\begin{align}\label{MaB4}
&\frac{1}{2\pi}\sum_{b=1}^d \frac{\partial^2\mathcal{V}}{\partial\lambda_a\partial\lambda_b}\vol(B_4)
+\frac{b_1}{2\pi}\sum_{i=1}^{r+1}\frac{\partial^2\mathcal{V}}{\partial\lambda_a\partial b_i} \int_{B_4}F_i\wedge J_{B_{4}}\nn
&\qquad\qquad\qquad\qquad-\frac{b_1}{2\pi}\sum_{i_1,i_2=1}^{r+1}
\frac{\partial^2 \mathcal{V}}{\partial {b_{i_2}}\partial v^{i_1}_a}
\int_{B_4}\tfrac{1}{2}F_{i_1}\wedge F_{i_2}=\nu_{r+2} M_a\,,
\end{align}
with $M_a\in\mathbb{Z}$.
Again, the $S_a$ are not linearly independent cycles in the fibre: multiplying 
(\ref{MaB4}) by $v_a^i$ and summing over $a=1,\ldots,d$, and using (\ref{constraintB4}), (\ref{NalphaB4}) and the identity (\ref{keyvRrel}), one 
can show that
\begin{align}\label{Mandnidentb4}
\sum_{a=1}^d v_a^i M_a = - \sum_{\alpha} N_\alpha n_i^\alpha\, ,
\end{align}
where the twisting parameters $n_i^\alpha$ were introduced in (\ref{Fi}). 
In the above expressions, from \eqref{b1form}, we should take
\begin{align}
b_1=\frac{2}{r}\,,
\end{align}
after taking derivatives with respect to $b_i$.
Finally, the supersymmetric action \eqref{susyact} can be written as
\begin{align}\label{SsusyB4}
\Ssusy
&=\nu_{r+2}\frac{2\pi }{r+1}\left(
\frac{1}{2\pi} a^\alpha N_\alpha-\sum_{a=1}^d\lambda_a {M_a}
\right)\,,
\end{align}
where recall that the $a^\alpha$ were introduced in \eqref{aalphadef} and parametrize the K\"ahler form, $J_{B_{2\Bp}}$, on the base $B_4$.

\subsection{$\X_{2r+1} \ \hookrightarrow \ Y_{2r+7} \ \rightarrow  B_6$}\label{beesix}

Now $r=0$ is relevant\footnote{{When $r=0$ the fibre $\X_1$ is simply a circle:
we have only one toric vector $v_1=1$, no K\"ahler parameters $\lambda_a$, and the master volume is simply $\mathcal{V}=2\pi/b_1$. 
There is only one twisting $U(1)$ bundle with curvature $F_1$, and moreover from 
(\ref{twist1}) we have $[F_1]=[\rho]$. In this case, the only terms which contribute are those involving only derivatives of $\mathcal{V}$ with respect to $b_1$. The formulae in this subsection then simply give rise to
the formulae \eqref{constraint}--\eqref{quantize} for the case of a regular $U(1)$ fibration
over $B_6$ with $\eta=\frac{1}{2}(\diff\varphi_1+P)$.}
} 
for the type IIB case while $r=1$ is relevant for the $D=11$ case.

The topological constraint condition \eqref{constraint} is given by
\begin{align}\label{conb6}
&\sum_{a,b=1}^d  \frac{\partial^2\mathcal{V}}{\partial\lambda_a\partial\lambda_b} \vol(B_6)
+b_1\sum_{i=1}^{r+1}\sum_{a=1}^d \frac{\partial^2\mathcal{V}}{\partial\lambda_a\partial b_i}\int_{B_6}F_i\wedge \tfrac{1}{2}J_{B_{6}}^2
-\sum_{a=1}^d \frac{\partial\mathcal{V}}{\partial\lambda_a}\int_{B_6}F_1\wedge \tfrac{1}{2}J_{B_{6}}^2\nn
&\qquad+{b_1^2}\sum_{i,j=1}^{r+1}\frac{\partial^2 \mathcal{V}}{\partial b_i\partial b_j}\int_{B_6}\tfrac{1}{2}F_i\wedge F_j\wedge J_{B_{6}}\nn
&\qquad-b_1^2\sum_{i_1,i_2,i_3=1}^{r+1}\frac{\partial^2}{\partial b_{i_2}\partial b_{i_3}}
\left(
\frac{1}{r+1}\sum_{a=1}^d\lambda^{a}\frac{\partial \mathcal{V}}{\partial v^{i_1}_a}
\right)
\int_{B_6}\tfrac{1}{3!}F_{i_1}\wedge F_{i_2}\wedge F_{i_3}=0
\,.
\end{align}
There are two types of flux integrals, corresponding to two types of $(2r+5)$-cycles. 
The first type of cycles have the fibred form 
\begin{align}
\X_{2r+1} \ \hookrightarrow \ \Sigma_\alpha \ \rightarrow C^{(4)}_\alpha\, ,
\end{align}
 with $C^{(4)}_\alpha\subset B_6$ a four-cycle. We find
\begin{align}\label{prefflux11}
&-\sum_{a=1}^d  \frac{\partial  \mathcal{V}}{\partial \lambda_a } \int_{C^{(4)}_\alpha}\tfrac{1}{2}J_{B_{6}}^2
-b_1\sum_{i=1}^{r+1}\frac{\partial  \mathcal{V}}{\partial b_i }\int_{C^{(4)}_\alpha}F_i\wedge J_{B_{6}}\nn
&+\frac{b_1}{r+1}\sum_{a=1}^{d}\sum_{i_1,i_2=1}^{r+1}
\lambda^{a}\frac{\partial^2 \mathcal{V}}{\partial b_{i_2}\partial v^{i_1}_{a}}\int_{C^{(4)}_\alpha}\tfrac{1}{2}F_{i_1}\wedge F_{i_2}
=\nu_{r+3} N_\alpha\,,
\end{align}
where $\nu_{r+3}$ is a non-zero, real constant,  fixed for the case of $r=0,1$ as in \eqref{nu3}, \eqref{nu4}, respectively, and $N_\alpha\in\mathbb{Z}$.
The second set of $(2r+5)$-cycles are given by
\begin{align}
S_a \ \hookrightarrow \ \Sigma_a \ \rightarrow  B_6\, ,
\end{align}
where $S_a$ is a $(2r-1)$-cycle in $\X_{2r+1}$ associated with a divisor on the cone $C(\X_{2r+1})$.
We find
\begin{align}
&\frac{1}{2\pi}\sum_{b=1}^d \frac{\partial^2\mathcal{V}}{\partial\lambda_a\partial\lambda_b}\vol(B_6)
+\frac{b_1}{2\pi}\sum_{i=1}^{r+1}\frac{\partial^2\mathcal{V}}{\partial\lambda_a\partial b_i} \int_{B_6}F_i\wedge \tfrac{1}{2}J_{B_{6}}^2\nn
&\qquad
-\frac{b_1}{2\pi}\sum_{i_1,i_2=1}^{r+1}
\frac{\partial^2 \mathcal{V}}{\partial b_{i_2}\partial v^{i_1}_{a}}
\int_{B_6}\tfrac{1}{2}F_{i_1}\wedge F_{i_2}\wedge J_{B_{6}}  \\
&\qquad
+\frac{b_1}{2\pi}\sum_{i_1,i_2,i_3=1}^{r+1}
\frac{\partial}{\partial b_{i_3}}
\left(\frac{1}{ (r+1)}\sum_{b_1=1}^d\lambda^{b_1} \frac{\partial^{2} \mathcal{V}}{\partial v^{i_1}_{b_1}\partial v^{i_2}_{a}}
\right)
\int_{B_6}\tfrac{1}{3!}F_{i_1}\wedge F_{i_2}\wedge F_{i_3}=\nu_{r+3} M_a\nonumber
\,,
\end{align}
with $M_a\in\mathbb{Z}$.
Again, the $S_a$ are not linearly independent cycles in the fibre: multiplying 
(\ref{MaB4}) by $v_a^i$ and summing over $a=1,\ldots,d$, and using \eqref{conb6},
\eqref{prefflux11}
one can show that
\begin{align}
\sum_{a=1}^d v_a^i M_a = - \sum_{\alpha} N_\alpha n_i^\alpha\, ,
\end{align}
where the $n_i^\alpha$ were introduced in (\ref{Fi}). In proving this, we have also used the identity \eqref{4indexidentity} in Appendix \ref{app:identities}.
In the above expressions, from \eqref{b1form}, we should take
\begin{align}
b_1=\frac{2}{r+1}\,,
\end{align}
after taking derivatives with respect to $b_i$.
Finally, the supersymmetric action \eqref{susyact} can be written as
\begin{align}\label{b6ssusy}
\Ssusy  =
\nu_{r+3}\frac{2\pi }{r+2}\left(
\frac{1}{2\pi} a^\alpha N_\alpha-\sum_{a=1}^d\lambda_a {M_a}
\right)\, ,
\end{align}
where the $a^\alpha$ were introduced in \eqref{aalphadef} 
and parametrize the K\"ahler form, $J_{B_{6}}$, on the base $B_6$.


\section{Examples}\label{sec:examples}

In this section we illustrate our general formalism and procedure in a variety of examples, focusing on the cases where the base space $B_{2\Bp}$ has complex dimension $\Bp=2$ and $\Bp=3$.\footnote{The 
Riemann surface case 
$\Bp=1$ has already been treated extensively in \cite{Gauntlett:2018dpc, Gauntlett:2019roi, Hosseini:2019ddy, Hosseini:2019use}.}  In addition to reproducing the results of some known explicit supergravity solutions 
summarized in Appendix \ref{app:examples}, where the bases $B_4$ and $B_6$ are K\"ahler-Einstein manifolds,  
we also work out examples where the base manifold is K\"ahler, but not Einstein. 
 In particular, we present
the calculations for $B_4=\Sigma_{g_1}\times \Sigma_{g_2}$, namely the product of two Riemann surfaces of genus $g_1$ and $g_2$, as well as for  $B_4=\mathbb{F}_n$,  the $n$th Hirzebruch surface. 

\subsection{Type IIB}\label{IIBcase}

In this subsection we consider $AdS_3\times Y_7$ examples of the form $\X_{3} \ \hookrightarrow \ Y_{7} \ \rightarrow  B_4$, for a variety of 
K\"ahler cases $B_4$. 

\subsubsection{$B_4=dP_k$}\label{sec:dPk}

We begin with the case that $B_4=dP_k$, the $k$th del Pezzo surface.\footnote{In the rest of the paper the base space has been denoted $B_{2\Bp}$, of complex dimension 
$\Bp$. In this section this $\Bp=2$, and in an abuse of notation instead in this subsection the integer $k=0,\ldots,8$ will label the del Pezzo surface $dP_k$.}
By definition this is the complex projective space $\mathbb{C}P^2$ blown up at $k=0,\ldots,8$ generic points. We let
\begin{align}
c_1 = 3H - \sum_{i=1}^kE_i
\end{align}
denote the anti-canonical class, where $H$ is the hyperplane class, and $E_i$ denote the exceptional divisors in the blow-up. 
We denote $M^{(k)}\equiv \int_{dP_k} c_1\wedge c_1  = 9-k$. 

For simplicity we will here only present the special case where the cohomology classes of the K\"ahler form 
 $J_{B_4}$ and curvatures of the fibration $F_i$ in $H^2(dP_k,\R)$ are proportional to the class $c_1$. 
We thus write
\begin{align}\label{classesdPk}
[J_{B_4}] = A \frac{c_1}{m_k}\, , \qquad \frac{1}{2\pi}[F_i] = n_i \frac{c_1}{m_k}\, ,
\end{align}
where $i=1,2$ and $A\in\R$, $n_i\in\Z$. Here $m_k$ is the 
\emph{Fano index}  of $dP_k$. By definition this is the largest positive 
integer so that $c_1/m_k\in H^2(dP_k,\Z)$ is an integer class. 
This is $m_0=3$ for $\mathbb{C}P^2$, but $m_k=1$ for the remaining del Pezzo surfaces $dP_k$ with $k=1,\ldots,8$. 
Furthermore, we note from  \eqref{twist1} that we have $n_1=m_k$.  
This case 
then has three flux quantum numbers: $N$, $M_i$, for $i=1,2$. In particular the two-cycle 
$C^{(2)}$ for flux quantum number $N$ in \eqref{NalphaB4}
is taken to be the Poincar\'e dual to $c_1/m_k$.

We first solve the constraint equation \eqref{constraintB4} for $A$ to obtain
\begin{align}
A=\frac{2 \pi  \left(  n_2^2 \lambda_1(2 p-b_2)^3+b_2^3 \lambda_2 (m_k p-n_2)^2\right)}{b_2 p (2 p-b_2) (2 n_2 (p-b_2)+m_k b_2 p)}\,, 
\end{align}
where here, as below, we have set $b_1=2$ as required for an $AdS_3$ solution after taking derivatives with respect to the $b_i$.
We then solve the expression for the preferred flux, $N$, in \eqref{NalphaB4}
for one of the transverse K\"ahler class parameters $\lambda_a$, specifically $\lambda_1$, to obtain
\begin{align}
\lambda_1= \frac{b_2 \lambda_2}{b_2-2 p}-\frac{m_k^2b_2  [2 n_2 (p-b_2)+m_k b_2 p]}{16 \pi ^3 M^{(k)} n_2 (m_k p-n_2)}\nu_3 N\, .
\end{align}
We then find that the two remaining fluxes can be expressed as
\begin{align}
M_1=\frac{(n_2- m_k p)}{p}N,\quad M_2=-\frac{n_2}{p}N\, ,
\end{align}
while the off-shell trial central charge function, given by \eqref{cS} and \eqref{SsusyB4},
takes the form
\begin{align}
\cZ=-\frac{3 m_k^2\left[n_2^2 \left(4 p^2-6 b_2 p+3 b_2^2\right)+m_k b_2 n_2 p (2 p-3 b_2)+m_k^2 b_2^2 p^2\right]}{M^{(k)} n_2 p (m_k p-n_2)}N^2\,.
\end{align}
Extremizing $\cZ$ over $b_2$ we find
\begin{align}
b_2=\frac{n_2 p (3 n_2- m_k p)}{m_k^2 p^2-3m_k n_2 p+3 n_2^2}\, ,
\end{align}
and hence $\csugra\equiv \cZ|_{\text{on-shell}}$ is given by
\begin{align}\label{fincciib}
 \csugra=\frac{9 m_k^2 n_2p (n_2- m_k p)}{M^{(k)} \left(m_k^2 p^2-3 m_k n_2 p+3 n_2^2\right)}
N^2\, .
\end{align}

We can now compare with the explicit $AdS_3\times \mathscr{Y}^{\pJ,\qJ}(KE_4^+)$
supergravity solutions of \cite{Gauntlett:2006af}, which are briefly
summarized in appendix \ref{iibexpl}. For each choice of $KE_4^+$ these
solutions are specified by two positive, relatively prime integers $\pJ>0,\qJ>0$, as
well as an overall flux number $n$. We make the obvious identifications $M=M^{(k)}$ and $m=m_k$, together with
$\pJ=-n_2$, $\qJ=p$ and $n= (m_k^2h/p M^{(k)}) N$, where $h\equiv \mathrm{hcf}(M^{(k)}/m_k^2,p)$.
We then notice that the flux quantum numbers can be written as
\begin{align}\label{emmandenn}
M_1 = (n_2-m_kp)\frac{M^{(k)}}{h m_k^2}n\, , \quad M_2 = - n_2 \frac{M^{(k)}}{h m_k^2}n\, , \quad N =  \frac{M^{(k)}}{m_k^2}\frac{p}{h} n\, .
\end{align}
Each term in the products is manifestly an integer, provided that $n$ is an integer, ensuring that $M_1,M_2,N\in \Z$.  Moreover, this ensures that 
\emph{all} flux quantum numbers are integer. To see this, recall from \eqref{classesdPk} that $[J_{B_4}]$ and $[F_i]$ are both proportional 
to the class $c_1$. From the general flux quantization condition \eqref{NalphaB4}, recalling that
$N$ is the flux through the two-cycle Poincar\'e dual to $c_1/m_k$ as well as the expression for $N$ in \eqref{emmandenn},
we may then deduce that the flux associated to an arbitrary 
two-cycle $C^{(2)}_\alpha\subset dP_k$ is 
\begin{align}
N_\alpha = \frac{p}{h}n \int_{C_\alpha^{(2)}} \frac{c_1}{m_k}\in \Z\, .
\end{align}
All flux quantum numbers are hence integer, provided that $n$ is an (arbitrary) integer.
We find that the fluxes of the explicit supergravity solutions, summarized in
\eqref{finalNIIB}, are related to the $M_i$ via $N(D_0)=M_1$ and $N(\tilde D_0)= -M_2$. 
Finally, the expression for the central charge, given in \eqref{fincciib} precisely agrees with
the expression obtained from the explicit supergravity solution \eqref{iibexplccfin}. 
We shall return to comment on the formula \eqref{fincciib} for $k=1,2$, where 
no K\"ahler-Einstein metric exists, in subsection \ref{sec:Fn}. 

\subsubsection{$B_4=\Sigma_{g_1}\times \Sigma_{g_2}$}\label{sec:prod}

We next examine the case when $B_4=\Sigma_{g_1}\times \Sigma_{g_2}$ is a product of two Riemann surfaces of genus $g_1$ and $g_2$. 
We introduce the normalized volume form classes $\vol_1$, $\vol_2$ for each Riemann surface, respectively, where 
$\int_{\Sigma_{g_1}}\vol_1 = 1 = \int_{\Sigma_{g_2}}\vol_2$. We may then write
\begin{align}
[J_{B_4}] = A_1\vol_1 + A_2\vol_2\, , \qquad \frac{1}{2\pi}[F_i] = n_i \vol_1 + k_i\vol_2\, ,
\end{align}
where $A_1,A_2\in \R$ and $n_i,k_i\in\Z$, $i=1,2$. We note from  \eqref{twist1} that we have $n_1=2-2g_1$, $k_1=2-2g_2$.  
This case 
then has four flux quantum numbers: $N_i$, $M_i$, for $i=1,2$. 

As in the previous subsection we first solve the constraint equation \eqref{constraintB4}, where we choose to eliminate the K\"ahler class parameter $A_1$. 
We then solve the expression for the fluxes $N_i$ given by \eqref{NalphaB4}, where $i=1,2$ labels the Riemann surfaces, 
and eliminate $A_2$ and $\lambda_1$. We then find that two remaining fluxes can be expressed as
\begin{align}
M_1 = \frac{[2 (g_1-1) p+n_2]N_2+ [2 (g_2 -1) p+k_2]N_1}{p}\, , \qquad M_2 = -\frac{k_2 N_1+n_2 N_2}{p}\, ,
\end{align}
while the off-shell trial entral charge function, given by \eqref{cS} and \eqref{SsusyB4},
may be computed as a function of $b_2$, and depends on the parameters $p,g_1,g_2,n_2,k_2,N_1,N_2$.
This may then be extremized over $b_2$ and evaluated on-shell. Rather than give the general expressions, which are rather unwieldy,
 we here present the special symmetric case where we choose $k_2=n_2\equiv k$ 
and $N_2=N_1\equiv N$. 
We then find that the extremal value of $b_2$ is given by
\begin{align}
b_2 = \frac{k p \left[p (g_1+g_2-2)+3 k\right]}{3 k p (g_1+g_2-2)+p^2 (g_1+g_2-2)^2+3 k^2}\, ,
\end{align}
while the central charge is
\begin{align}\label{cg1g2}
\csugra& = 
\frac{18 k  p \left[p (g_1+g_2-2)+k\right]}{3 k p (g_1+g_2-2)+p^2 (g_1+g_2-2)^2+3 k^2}N^2\, . 
\end{align}

Setting $g_1=g_2=0$, which corresponds to $B_4=S^2\times S^2$, we find that the 
central charge \eqref{cg1g2} matches the explicit supergravity solution result in \eqref{iibexplccfin}, where 
$m=2$, $M=8$, and we identify parameters as $\pJ=-k$, $\qJ=p$ and $n=(h/p)N$. In particular we also then 
find that the fluxes of the explicit supergravity solutions, summarized in
\eqref{finalNIIB}, are related to the $M_i$ via $N(D_0)=M_1$ and $N(\tilde D_0)= -M_2$.

\subsubsection{$B_4=\mathbb{F}_n$}\label{sec:Fn}

Finally, we examine the case when $B_4=\mathbb{F}_n$ is the $n$th Hirzebruch surface. This is the complex surface 
defined as the total space of a $\mathbb{C}P^1$ bundle over $\mathbb{C}P^1$. There are various equivalent ways to describe 
the fibration. For example, one can take the complex line bundle $\mathcal{O}(-n)$ over $\mathbb{C}P^1$, and 
add a point at infinity to each fibre to make the fibres Riemann spheres $\mathbb{C}\cup \{\infty\}\cong \mathbb{C}P^1$. Alternatively, one can take the projectivization 
$\mathbb{P}(\mathcal{O}(0)\oplus \mathcal{O}(-n))$. In the first description, we shall refer to the origin of the complex line fibre 
as the south pole of the Riemann sphere, and the point at infinity that we add as the north pole. These give rise to sections 
$S_1$, $S_2$ of the $\mathbb{C}P^1$ bundle over $\mathbb{C}P^1$, respectively, which have intersection numbers 
$S_1\cdot S_1=n$, $S_2\cdot S_2 = -n$, $S_1\cdot S_2=0$. Another natural two-cycle is the class $F$ of the fibre, at a fixed point 
on the $\mathbb{C}P^1$ base. This clearly has intersection numbers $F\cdot S_1=1=F\cdot S_2$, $F\cdot F=0$. 
A convenient basis of two-cycles for $H_2(\mathbb{F}_n,\Z)\cong \Z^2$ is then $\{F,S_1\}$. With respect to this basis, 
the above formulae imply that the intersection form is
\begin{align}
I  = \left(\begin{array}{cc} 0 & 1 \\ 1 & n\end{array}\right)\, .
\end{align}
We denote the Poincar\'e dual two-form basis for $\{F,S_1\}$ as $\{e_1=\hat{F},e_2=\hat{S}_1\}$. These form a dual basis 
for the cohomology $H^2(\mathbb{F}_n,\Z)\cong \Z^2$, where 
\begin{align}
\int_{\mathbb{F}_n} e_\alpha\wedge e_\beta = I_{\alpha\beta}\, , \qquad\qquad \alpha,\beta=1,2\, .
\end{align}

With this notation in hand, we may then write the 
cohomology classes of the K\"ahler form $J_{B_4}$ and curvature two-forms $F_i$ in $H^2(\mathbb{F}_n,\R)$ as
\begin{align}\label{ffifteen}
[J_{B_4}] = \sum_{\alpha=1}^2 A_\alpha e_\alpha\, , \qquad \frac{1}{2\pi}[F_i] = \n_i e_1 + \m_i e_2\, ,
\end{align}
where $A_\alpha\in \R$, $\n_i,\m_i\in \Z$ and $i=1,2$, $\alpha=1,2$. The anti-canonical class of $\mathbb{F}_n$ 
is given by
\begin{align}
c_1(\mathbb{F}_n)= 2\hat{F}+\hat{S}_1+\hat{S}_2 = (2-n)\hat{F} + 2 \hat{S}_1 = (2-n)e_1 + 2e_2\, ,
\end{align}
where in the second step we have used the fact that $nF = S_1-S_2$, with the same linear relation of course holding for the 
Poincar\'e duals. From equation \eqref{twist1} we thus deduce that 
\begin{align}\label{Fncan}
\n_1 = 2-n\, , \qquad \m_1 = 2\, .
\end{align}

As in the previous subsection we first solve the constraint equation \eqref{constraintB4}, where we choose to eliminate the K\"ahler class parameter $A_1$. 
We then solve the expression for the fluxes $N_\alpha$ given by \eqref{NalphaB4}, where $\alpha=1,2$ labels the basis two-cycles in $\mathbb{F}_n$, 
and eliminate $A_2$ and $\lambda_1$. We then find that two remaining fluxes can be expressed as
\begin{align}
M_1 = \frac{N_1 [(n-2) p+\n_2]+N_2 (\m_2-2 p)}{p}\, , \qquad M_2 =-\frac{\n_2 N_1 + \m_2N_2}{p}\, .
\end{align}
The off-shell trial entral charge function, given by \eqref{cS} and \eqref{SsusyB4},
may be computed as a function of $b_2$, and depends on the parameters $p,n,\n_2,\m_2,N_1,N_2$.
This may then be extremized over $b_2$ and evaluated on-shell, to obtain the central charge
\begin{align}\label{Fnc}
\csugra& = 3 N_1 p (n N_1-2N_2) \Big\{\m_2 \Big[n^2 N_1^2 (p-\m_2)+2 n N_1(N_2 (\m_2-3 p)+N_1 p) \nn 
& +4 N_2 (N_1 p+(2p-\m_2) N_2 )\Big]-2 \n_2 N_1(\m_2 n N_1+2N_2 (\m_2-p)+(n-4) N_1 p)\nn 
& -4 \n_2^2 N_1^2\Big\}\Big/\Big\{N_1^2 \Big[\m_2^2 n^2-\m_2 n (n+2) p+2 \m_2 n \n_2+(n-2)^2 p^2+2 (n-4) \n_2 p \nn 
& +4 \n_2^2\Big]-2 N_1 N_2 (2 p-\m_2) ((n-2) p-\m_2 n)+4 \n_2 N_1N_2 (\m_2-p)\nn 
& +4N_2^2 (p-\m_2)^2\Big\}\, .
\end{align}

We may compare certain special cases of the  rather unwieldy general result \eqref{Fnc} with our earlier results. 
First, taking $n=0$ gives the product base $B_4=\mathbb{C}P^1\times \mathbb{C}P^1 = S^2\times S^2$, which is the 
genus $g_1=0=g_2$ case from subsection \ref{sec:prod}. Further specializing to the symmetric case where we choose 
$\n_2=\m_2\equiv k$, $N_1=N_2\equiv N$, we find that \eqref{Fnc} agrees precisely with the product of 
Riemann surfaces central charge \eqref{cg1g2} with $g_1=0=g_2$, as it should do. 

Secondly, $\mathbb{F}_1=dP_1$. Comparing to the notation of subsection \ref{sec:dPk}, the north and south pole 
sections of $\mathbb{F}_1$ then have Poincar\'e duals $\hat{S}_1=H$, $\hat{S}_2=E_1$, where recall that $H$ is the hyperplane class and $E_1$ is the exceptional divisor class. The basis $\{e_1=\hat{F},e_2=\hat{S}_1\}$ we have used in this subsection is hence $\{e_1=H-E_1,e_2=H\}$. Moreover, the quantum number $N$ in subsection \ref{sec:dPk} is by definition the flux through the Poincar\'e dual of the 
anti-canonical class $3H-E_1=e_1+2e_2$, implying that $N=N_1+2N_2$. Moreover, since in \eqref{classesdPk} both 
$[J_{B_4}]$ and $[F_i]$ are proportional to the same class $c_1=3H-E_1=e_1+2e_2$, 
from the expression \eqref{NalphaB4} for the flux quantum numbers $\{N_1,N_2\}$, defined to be the flux through the Poincar\'e duals 
to $\{e_1,e_2\}$, respectively, we deduce that
\begin{align}
\begin{pmatrix} N_1 \\ N_2 \end{pmatrix} \propto I \cdot \begin{pmatrix} 1 \\ 2 \end{pmatrix} = \begin{pmatrix} 2 \\ 3 \end{pmatrix}\, .
\end{align} 
Thus $N_2=3N_1/2$. Combining this with $N=N_1+2N_2$ above we deduce that $N_1=N/4$, $N_2=3N/8$. 
Finally, the $U(1)$ flavour twisting in \eqref{classesdPk} satisfies $\tfrac{1}{2\pi}[F_2]=n_2 c_1$ 
and comparing with \eqref{ffifteen}, \eqref{Fncan} we can identify
\begin{align}
\n_2 = (2-n)n_2 = n_2\, \qquad \m_2 = 2n_2\, ,
\end{align}
where we set $n=1$ for the first Hirzebruch surface $\mathbb{F}_1$. 
Making these substitutions in \eqref{Fnc} one finds
\begin{align}
\csugra&= \frac{9 n_2 p (n_2-p)}{8 \left(p^2-3 n_2 p+3 n_2^2\right)}N^2\, .
\end{align}
This agrees with the central charge \eqref{fincciib} for $dP_1$ on setting $m_1=1$, $M^{(1)}=9-1=8$ for the the first del Pezzo surface. 
Of course, in this case $dP_1$ does not admit a K\"ahler-Einstein metric, and so we cannot compare with 
the explicit supergravity solution result \eqref{iibexplccfin}. However, it is natural to conjecture that a corresponding 
GK supergravity solution \emph{does} exist in this case, but simply outside the K\"ahler-Einstein ansatz 
utilized in \cite{Gauntlett:2006af}. Similar remarks apply to the central charge \eqref{fincciib} for second del Pezzo surface $k=2$, 
which is also not K\"ahler-Einstein. Whether GK supergravity solutions exist for general Hirzebruch surfaces $\mathbb{F}_n$, for which the central 
charge is then given by \eqref{Fnc}, is an interesting 
open problem. 

\subsection{$D=11$}\label{d11exsolcase}

In this subsection we consider an $AdS_2\times Y_9$ example of the form $\X_{3} \ \hookrightarrow \ Y_{9} \ \rightarrow  B_6$, where we take the K\"ahler base to be $B_6=\mathbb{C}P^3$

We let $H$ generate  the second cohomology $H=1\in H^2(\mathbb{C}P^3,\Z)\cong \Z$ of $\mathbb{C}P^3$, which satisfies
$\int_{\mathbb{C}P^3} H^3 =1$. We may then write the cohomology classes of the K\"ahler form $J_{B_6}$ and curvature two-forms $F_i$ in $H^2(\mathbb{C}P^3,\R)$ as
\begin{align}
[J_{B_6}] = A H \, , \qquad \frac{1}{2\pi}[F_i] = n_i H\, ,
\end{align}
where $i=1,2$ and $A\in\R$, $n_i\in\Z$. 
Furthermore, from \eqref{twist1} we then have $n_1=4$.
This case 
has three flux quantum numbers: $N$, $M_i$, for $i=1,2$.
In carrying out our general procedure we will see that some ambiguities arise. We believe that it should
be possible to fix these ambiguities by imposing suitable positivity conditions on the K\"ahler class parameters $A$ and $\lambda_a$, but we leave a general discussion of this for future work. Here we are content to
show that
there is a solution that gives precisely the same value for the entropy as that obtained from the explicit supergravity solutions discussed in Appendix \ref{appdd11}.

We first solve the constraint equation \eqref{conb6} for $A$, finding two solutions. In continuing the procedure, we find that one of these solutions ultimately gives rise to an action function that, after setting $b_1=1$ as required for an $AdS_2$ solution, only depends linearly on $b_2$ and hence we cannot solve for $b_2$ after
extremizing this action. We thus continue with the other solution for $A$ which, with $b_1=1$, and $n_1=4$, is given by
\begin{align}
A=
\frac{2 \pi  (4 b_2-n_2) \left[b_2^2 \lambda_2 (4 p-n_2)-\lambda_1 n_2 (b_2-p)^2\right]}{b_2 (p-b_2) (-2 b_2 n_2+4 b_2 p+n_2 p)}\,.
\end{align}

We next solve the expression for the preferred flux $N$, given in \eqref{prefflux11}, for $\lambda_1$, again
finding two solutions. These are rather lengthy and we do not record them here. 
For both solutions
the remaining two fluxes take the form
\begin{align}
M_1=\frac{(n_2-4 p)}{p}N,\qquad M_2=-\frac{n_2}{p}N\, ,
\end{align}
which we note implies that $N$ is divisible by $p$.
Furthermore, the two solutions for $\lambda_1$ just give rise to a change in sign of
the action. We find that one of these solutions, which we now continue with, 
leads to precisely the entropy of the explicit solutions in Appendix \ref{appdd11}. We then obtain an expression for
the off-shell entropy function, and may set $b_1=1$. After varying 
with respect to the remaining $R$-symmetry direction $b_2$ we find that there are two extremal values for $b_2$, one of which connects
with the explicit supergravity solutions. Assuming $p>0$ and $n_2<0$, which we will see in a moment are conditions imposed from the explicit  solutions in Appendix \ref{appdd11}, we find that this
specific solution for $b_2$ is given by
\begin{align}
b_2=\frac{n_2 \left(\sqrt{-4 n_2 p+n_2^2+8 p^2}+n_2-4 p\right)}{4 \sqrt{-4 n_2 p+n_2^2+8 p^2}}\,.
\end{align}
Furthermore, with this value of $b_2$, the on-shell entropy $\mathscr{S}$, given by
\eqref{cS2} and \eqref{b6ssusy},
takes the form
\begin{align}\label{d11entbulk}
\mathscr{S}=\frac{\sqrt{2} \pi  \left(\sqrt{-4 n_2 p+n_2^2+8 p^2}+n_2-2 p\right) \sqrt{n_2(n_2-4 p)}}{3 p^{3/2}}N^{3/2}\,.
\end{align}

Having obtained this result from our general procedure, we may now compare with the explicit $AdS_2\times \mathscr{Y}^{\pJ,\qJ}(KE_6^+)$
solutions of $D=11$ supergravity
that are discussed in Appendix \ref{appdd11}. These solutions are 
labelled by two relatively prime integers $\pJ,\qJ>0$, as well as an overall flux number $n$. 
The parameters are related by
 $\pJ=-n_2$, $\qJ=p$ and $n=N/p$. Notice that these imply
$p>0$ and $n_2<0$ which we used above, and also that the identification on $n$ is consistent
with the fact that, as noted above, $p$ divides $N$. We then have that the fluxes of the $D=11$
solution given in \eqref{finalNsd11} are related to $M_1,M_2$ via
$N(D_0)=M_1$, $N(\tilde D_0)= -M_2$. Finally, one can also check that
the on-shell entropy given in \eqref{d11entbulk} precisely agrees with that given
in \eqref{finalentd11}.


\section{Discussion}\label{discsec}
 
We have studied the geometric extremal problem,
introduced in \cite{Gauntlett:2007ts}, for GK manifolds $Y_{2n+1}$, $n\ge 3$, 
that are toric fibrations over a K\"ahler base manifold $B_{2\Bp}$.
Our results extend those of \cite{Gauntlett:2018dpc,Gauntlett:2019roi}, which studied the cases
of $Y_7$ and $Y_9$ torically fibred over a Riemann surface  $B_2=\Sigma_g$, respectively. 
Similar to \cite{Gauntlett:2018dpc,Gauntlett:2019roi}, we have shown that the relevant
flux quantization conditions and the constraint condition, as well as the action function that determines the supersymmetric $R$-symmetry Killing vector, may all be written in terms of the master volume of the toric fibre, together with certain global data associated 
with the K\"ahler base.  We have also checked our new formulae using explicit classes of supergravity solutions
of the form $AdS_3\times Y_7$ and $AdS_2\times Y_9$, finding exact agreement.

When introducing the toric fibres our starting point was to consider them to be Sasaki. 
Such fibres have toric data, specified by a set of inward pointing normal vectors $\vec{v}_a$, 
that are associated with convex polyhedral cones. However, we know from explicit examples that this is too restrictive
and we should also allow vectors $\vec{v}_a$ that are associated with ``non-convex" toric cones, as introduced in \cite{Gauntlett:2018dpc}. 
Further study of such novel toric geometry is certainly warranted and this could also help to resolve
the ambiguities in carrying out the extremal problem that we saw in section \ref{d11exsolcase} for certain examples.
More generally, an important outstanding topic is to 
determine the necessary and sufficient conditions for the existence of the GK geometries, given the K\"ahler base and the toric fibre data. 

A natural way in which the $AdS_3\times Y_7$ and $AdS_2\times Y_9$ supergravity solutions arise is to consider wrapping a stack of D3-branes 
or membranes on a holomorphic curve in a Calabi-Yau four-fold or five-fold, respectively, as clarified in \cite{Gauntlett:2007ph,MacConamhna:2006nb}.
These configurations give rise to supersymmetric field theories in the unwrapped directions of the branes, and when these flow to a conformal fixed point, the supergravity dual develops an $AdS_3$ or $AdS_2$ factor, respectively. This perspective should be helpful in 
further understanding the geometries we have studied in this paper, as well as 
identifying the dual SCFTs.

\subsection*{Acknowledgments}
We would like to thank the Centro de Ciencias de Benasque Pedro Pascual for hospitality while part of this work was carried out. 
JPG is supported by the European Research Council under the European Union's Seventh Framework Programme (FP7/2007-2013), ERC Grant agreement ADG 339140. JPG is also supported by STFC grant ST/P000762/1, EPSRC grant EP/K034456/1, as a KIAS Scholar and as a Visiting Fellow at the Perimeter Institute. JPG acknowledges hospitality from the KITP and
support from National Science Foundation under Grant No. NSF PHY-1748958.
DM would like to thank the Galileo Galilei Institute for hospitality while this work was being carried out. 


\appendix

\section{Master volume identities}\label{app:identities}

In this appendix we derive a number of identities satisfied by the master volume $\mathcal{V}$, that are used in the main text. 

As described in section \ref{sec:master},
we begin by introducing the new coordinates $x_i\equiv y_i-y^{(0)}_i$, $i=1,\ldots,r+1$, on $\R^{r+1}$, so that we may write the master volume as
\begin{align}\label{mvappone}
\mathcal{V} = (2\pi)^{r+1}\int_{\R^{r+1}} \prod_{b=1}^d \theta((\vec{\x},\v_b)-\lambda_b) \delta((\vec{\x},\vec{b}))\, .
\end{align} 
Here the integration uses the standard Euclidean measure $\diff x_1\wedge\cdots\wedge \diff x_{r+1}$.
Using Stokes' theorem we have
\begin{align}\label{stokes}
\int_{\R^{r+1}} \nabla\left[f(\vec{x})\prod_{b=1}^d \theta((\vec{\x},\v_b)-\lambda_b) \delta((\vec{\x},\vec{b}))\right] = 0\, ,
\end{align}
where $f(\vec{x})$ is an arbitrary function. The boundary term at infinity here vanishes on integrating by parts, because the term in square brackets is 
compactly supported (on a compact polytope embedded in $\R^{r+1}$). Taking $f\equiv 1$ to be constant, and then 
computing the gradient, one obtains
\begin{align}\label{stokesstep}
& \sum_{a=1}^d \vec{v}_a \int_{\R^{r+1}}\delta((\vec{x},\vec{v}_a)-\lambda_a)\prod_{b\neq a} \theta((\vec{\x},\v_b)-\lambda_b) \delta((\vec{x},\vec{b}))\nonumber\\
& + \vec{b}\int_{\R^{r+1}}\prod_{b=1}^d \theta((\vec{\x},\v_b)-\lambda_b) \delta'((\vec{x},\vec{b}))=0\, .
\end{align}
Note immediately that the integral on the first line is proportional to $\partial\mathcal{V}/\partial\lambda_a$. We next 
 need to deal with the 
derivative of the $\delta$-function on the second line. Using the chain rule $\nabla = \vec{b}\, \partial_s$, where 
$s\equiv (\vec{x},\vec{b})$ is the argument of the $\delta$-function, and it follows that we may write
$\delta'(s) = \frac{\vec{b}\cdot \nabla}{|\vec{b}|^2} \delta(s)$. We then integrate the second line of 
(\ref{stokesstep}) by parts to obtain
\begin{align}
\sum_{a=1}^d \vec{v}_a \frac{\partial \mathcal{V}}{\partial \lambda_a} + (2\pi)^{r+1}\vec{b}\sum_{a=1}^d \frac{\vec{b}\cdot \vec{v}_a}{|\vec{b}|^2} 
\int_{\R^{r+1}}\delta((\vec{x},\vec{v}_a)-\lambda_a)\prod_{b\neq a} \theta((\vec{\x},\v_b)-\lambda_b) \delta((\vec{x},\vec{b}))=0\, ,\nonumber
\end{align}
which immediately gives
\begin{align}\label{NEWidentity}
\sum_{a=1}^d \left(\vec{v}_a-\vec{b}\, \frac{\vec{b}\cdot\vec{v}_a}{|\vec{b}|^2}\right)\frac{\partial\mathcal{V}}{\partial\lambda_a}=0\, .
\end{align}
Notice this identity holds for arbitrary vectors $\{\vec{v}_a\}$ and $\vec{b}$. 
When $v_a^1=1$ for all $a=1,\ldots,d$, which is true when $C(\X_{2r+1})$ is Gorenstein, one 
may take the $i=1$ component of (\ref{NEWidentity}), and substituting this back in one immediately derives
\begin{align}
\sum_{a=1}^d \left(\vec{v}_a-\frac{\vec{b}}{b_1}\right)\frac{\partial\mathcal{V}}{\partial\lambda_a}=0\, ,
\end{align}
which is equation (\ref{keyvRrel}) in the main text.

We next compute, from \eqref{mvappone}, that
\begin{align}
\frac{\partial\mathcal{V}}{\partial b_i} = (2\pi)^{r+1}\int_{\R^{r+1}} \prod_{b=1}^d \theta((\vec{\x},\v_b)-\lambda_b) \delta'((\vec{x},\vec{b}))x^i\, .
\end{align}
Integrating the $\delta$-function by parts, as we did above, one finds
\begin{align}
\frac{\partial\mathcal{V}}{\partial b_i} = & -(2\pi)^{r+1}\sum_{a=1}^d \frac{\vec{b}\cdot\vec{v}_a}{|\vec{b}|^2} \int_{\R^{r+1}} \delta((\vec{x},\vec{v}_a)-\lambda_a) \prod_{b\neq a} \theta((\vec{\x},\v_b)-\lambda_b) \delta((\vec{x},\vec{b}))x^i\nonumber\\
&-(2\pi)^{r+1}\frac{b^i}{|\vec{b}|^2}\int_{\R^{r+1}} \prod_{b=1}^d \theta((\vec{\x},\v_b)-\lambda_b) \delta((\vec{x},\vec{b}))\, .
\end{align}
But this simply reads
\begin{align}\label{Anotheridentity}
\frac{\partial \mathcal{V}}{\partial b_i} = -\sum_{a=1}^d \frac{\vec{b}\cdot\vec{v}_a}{|\vec{b}|^2}\frac{\partial\mathcal{V}}{\partial v_a^i} -\frac{b^i}{|\vec{b}|^2}\mathcal{V}\, .
\end{align}
Note that 
we can immediately deduce that
\begin{align}
\vec{b}\cdot \frac{\partial \mathcal{V}}{\partial \vec{v}_a} = 0\, , \qquad a=1,\ldots,d\, ,
\end{align}
by computing the expression for $\partial\mathcal{V}/\partial v_a^i$. Dotting (\ref{Anotheridentity}) 
with $\vec{b}$ then implies that $\mathcal{V}$ is homogeneous degree $-1$ in $b_i$.

Next taking $f=x^j$ in (\ref{stokes})
and computing in a similar way, we derive
\begin{align}\label{xjidentity}
 \delta_{ij} \mathcal{V} + \sum_{a=1}^d v_a^i \frac{\partial\mathcal{V}}{\partial v_a^j} - b_i\sum_{a=1}^d\frac{\vec{b}\cdot\vec{v}_a}{|\vec{b}|^2} \frac{\partial\mathcal{V}}{\partial v_a^j} - \frac{b_ib_j}{|\vec{b}|^2}\mathcal{V}=0\, .
\end{align}
Combining this with (\ref{Anotheridentity}) then gives the remarkably simple identity
\begin{align}\label{simpleidentity}
 \delta_{ij}\mathcal{V} + b_i\frac{\partial\mathcal{V}}{\partial b_j}+\sum_{a=1}^d v_a^i\frac{\partial\mathcal{V}}{\partial v_a^j}=0\, .
\end{align}
Contracting indices implies that for $\X_{2r+1}$, where recall $i,j=1,\dots,r+1$, we have
\begin{align}
\sum_{a=1}^d \vec{v}_a\cdot \frac{\partial\mathcal{V}}{\partial \vec{v}_a} +r\mathcal{V}= 0\, .
\end{align}
Differentiating (\ref{simpleidentity}) with respect to $b_k$ then immediately gives
\begin{align}\label{3indexproven}
 \delta_{ij}\frac{\partial \mathcal{V}}{\partial b_k} + \delta_{ik}\frac{\partial \mathcal{V}}{\partial b_j} + b_i\frac{\partial^2\mathcal{V}}{\partial b_j\partial b_k} + \sum_{a=1}^d v_a^i\frac{\partial^2\mathcal{V}}{\partial v_a^j\partial b_k}=0\, ,
\end{align}
or equivalently 
\begin{align}\label{3indexidentityagain}
 \Big[\frac{\partial\mathcal{V}}{\partial b_j}\delta_{ki}
+ \frac{b_i}{2}\frac{\partial^2\mathcal{V}}{\partial b_j\partial b_k} + \frac{1}{2}\sum_{a=1}^d v_a^i\frac{\partial^2\mathcal{V}}{\partial v_a^{j}\partial b_{k}}\Big]_{\mathrm{Sym}(j,k)}=0  \, ,
\end{align}
where ${\mathrm{Sym}(j,k)}$ denotes that we should symmetrize over the $j,k$ indices. Equation (\ref{3indexidentityagain}) is another identity used in  deriving results in the main text. 
Notice that we may compute
\begin{align}\label{deeveevee2}
\frac{\partial^2 \mathcal{V}}{\partial v^j_a \partial b_k}&=
{(2\pi)^{r+1}}\int_{\R^{r+1}} \delta((\vec{x},\v_a)-\lambda_a)\prod_{b\ne a} \theta((\vec{x},\v_b)-\lambda_b) \delta'((\vec{x},\vec{b}))x^jx^k\, ,
\end{align}
where the right hand side is manifestly symmetric in $j$ and $k$.
Finally, using (\ref{3indexproven}) one easily derives the identity
\begin{align}\label{4indexidentity}
& \Bigg[ \sum_{b=1}^d\lambda^b\frac{\partial}{\partial v^l_b}\left(
\frac{1}{2}\frac{\partial\mathcal{V}}{\partial b_j}\delta_{ki}
+ \frac{b_i}{3!}\frac{\partial^2\mathcal{V}}{\partial b_j\partial b_k} 
+ \frac{1}{3!}\sum_{a=1}^d v_a^i\frac{\partial^2\mathcal{V}}{\partial {v^a}_{j}\partial b_{k}}\right)\nonumber\\
& - \frac{1}{3!}\sum_{b=1}^d \lambda^b \frac{\partial^2\mathcal{V}}{\partial v^b_j\partial b_k}\delta_{li}\Bigg]_{\mathrm{Sym}(j,k,l)} = 0 \, ,
\end{align}
where ${\mathrm{Sym}(j,k,l)}$ denotes that we should symmetrize over the $j,k,l$ indices. Again, this is another identity 
needed to obtain results in the main text.

\section{Master volume for $\X_3$}\label{app:X3}

In this appendix we derive the formula (\ref{VX3}) for the master volume of $\X_3$.

As in section \ref{sec:fibre} we take the fibre $\X_3$ to be the link of a Gorenstein K\"ahler cone of complex dimension $r+1=2$. 
The toric data 
is then $v_1=(1,0)$, $v_2=(1,p)$, $p\in\mathbb{N}$, 
which describes an $A_{p-1}$ singularity, $C(\X_3)=\C^2/\Z_p$. Here the $\Z_p$ 
action on $\C^2$ is $(z_1,z_2)\mapsto (\omega_p z_1,\omega_p^{-1}z_2)$, where $\omega_p$ is a primitive $p$th root of unity. 
The outward pointing normals to the edges at the apex are $u_1=(0,1)$, $u_2=(p,-1)$. Recall also that the ``origin'' of the polytope $\mathcal{P}$ is located at
\bea
\vec{y}^{(0)} = \left(\frac{1}{2b_1},0\right)~.
\eea
Denoting by $\lambda_1$, $\lambda_2$ the K\"ahler parameters associated to the two facets, as in the main text, then the two vertices $\vec{y}_a$, $a=1,2$,
of the polytope $\mathcal{P}$, which here is a line segment, satisfy the equations
\bea
(\vec{y}_a-\vec{y}^{(0)},\vec{b}) = 0~, \qquad (\vec{y}_a-\vec{y}^{(0)},\vec{v}_a) = \lambda_a, \qquad a=1,2~.
\eea
These are easily solved to give
\bea
\left(\vec{y}_a-\vec{y}^{(0)}\right)_i = \frac{\lambda_a \varepsilon_{ij}b^j}{\varepsilon_{mn}v_a^mb^n}~.
\eea
The master volume (\ref{VEuc}) is 
\bea\label{linevertices}
\mathcal{V} =  \frac{(2\pi)^2}{|\vec{b}|}\vol(\mathcal{P})\,,
\eea
where here the ``volume'' of the polytope $\mathcal{P}$ is simply the length of the line segment between the two vertices in (\ref{linevertices}). But this is 
\bea
\vol(\mathcal{P}) = \left(-\frac{\lambda_1\varepsilon_{ij}b^j}{\varepsilon_{mn}v_1^m b^n} + \frac{\lambda_2\varepsilon_{ij}b^j}{\varepsilon_{mn}v_2^m b^n}\right)\frac{\varepsilon_{ik}b^k}{|\vec{b}|}~.
\eea
Here each of the two terms in the bracket is the vector from the origin to the corresponding vertex. To compute the length we have then taken a two-dimensional cross product with the unit 
normal $\vec{b}/|\vec{b}|$. A short computation then gives the simple formula 
\begin{align}\label{VolX3}
\mathcal{V} =  (2\pi)^2\sum_{a=1}^2(-1)^a \frac{\lambda_a}{[\vec{v}_a,\vec{b}]}
=(2\pi)^2\left(-\frac{\lambda_1}{b_2} +\frac{\lambda_2}{b_2-b_1 p}\right)\, ,
\end{align}
where here $[\vec{v}_a,\vec{b}]$ is the determinant of the $2\times 2$ matrix, i.e. $[\vec{v}_a,\vec{b}]\equiv  \varepsilon_{mn}v_a^m b^n$.

Notice that setting  $\lambda_a=-\frac{1}{2b_1}$ for $a=1,2$, and then $b_1=2$, the extremum of the corresponding volume function $\mathcal{V}$
occurs at $b_2=p$, with extremal volume $ \mathcal{V}=2\pi^2/p$. This is the correct volume of the Lens space $L(1,p)=S^3/\Z_p$, 
equipped with its Sasaki-Einstein metric.

\section{More on the master volume}\label{app:derivatives}

In section \ref{sec:fibredGK} we presented formulae for the constraint (\ref{constraint}), flux quantization conditions (\ref{quantize}) and supersymmetric 
action (\ref{susyact}) for fibred GK geometries, where crucially the formulas depend only on certain topological integrals over the K\"ahler base $B_{2\Bp}$, together 
with derivatives of the master volume $\mathcal{V}$ of the fibres $\X_{2r+1}$. Here $\mathcal{V}=\mathcal{V}(\vec{b};\{\lambda_a\};\{\vec{v}_a\})$ 
is a function of the $R$-symmetry vector $\vec{b}$, K\"ahler class parameters $\{\lambda_a\}$, and the toric data $\{\vec{v}_a\}$ of the cone $C(\X_{2r+1})$. 
Derivatives of $\mathcal{V}$ with respect to all three appear. In section \ref{app:der} we derive various formulae for these derivatives that we have used 
in deriving the results of sections \ref{beetwo}, \ref{beefour} and \ref{beesix}.
In section \ref{app:beefour} we explicitly explain how these have been used 
to derive the expressions in section \ref{beefour} -- the results of  sections~\ref{beetwo} and \ref{beesix} are obtained similarly.
 As we have  seen, there are various equivalent ways to write the master volume, 
and some forms are more useful than others in deriving particular formulae. We have already presented a number of identities satisfied by the master 
volume and its derivatives in Appendix \ref{app:identities}.

\subsection{Derivatives of the master volume}\label{app:der}

The master volume of the fibre $\X_{2r+1}$ can be written variously as
\begin{align}
\label{genvol}
\mathcal{V} & =  \int_{\X_{2r+1}}\eta\wedge \tfrac{1}{r!}\omega^r \nn&= \frac{(2\pi)^{r+1}}{|\vec{b}|}\vol (\mathcal{P})\nonumber \\
&  = {(2\pi)^{r+1}}\int_{\mathbb{R}^{r+1}}\prod_{a=1}^d \theta((\vec{x},\v_a)-\lambda_a) \delta((\vec{x},\vec{b}))\, .
\end{align}
The first equality is the original definition (\ref{defVmaaster}), while the second writes this in terms of the Euclidean volume of the moment 
map polytope $\mathcal{P}$, where the latter is defined in equation (\ref{generalP}), while the third equality writes this Euclidean volume 
in terms of step functions and a $\delta$-function integrated over $\R^{r+1}$. Recall here that $\R^{r+1}$ is parametrized by the shifted moment map 
variables $x_i$, $i=1,\ldots,r+1$, defined in (\ref{xi}), where we have suppressed  the standard Euclidean measure $\diff x_1\wedge \cdots\wedge \diff x_{r+1}$ 
in the notation, to keep formulae uncluttered. 

Using equation (\ref{SaVnew}) with $s=0$ we also have
\begin{align}\label{SaV}
-\frac{1}{2\pi} \frac{\partial \mathcal{V}}{\partial \lambda_a} & = \int_{S_a} \eta\wedge\tfrac{1}{(r-1)!}\omega^{r-1}\nonumber\\
&= {(2\pi)^r}\int_{\mathbb{R}^{r+1}}
\delta((\vec{x},\v_a)-\lambda_a)\prod_{b\ne a} \theta((\vec{x},\v_b)-\lambda_b) \delta((\vec{\x},\vec{b}))\, .
\end{align}
Recall here that $S_a\subset \X_{2r+1}$ is the $(2r-1)$-submanifold in $\X_{2r+1}$ associated with the $a$th toric divisor on 
the cone $C(\X_{2r+1})$, which moreover is Poincar\'e dual to $c_a$. Here the second line of (\ref{SaV}) follows immediately 
by differentiating the expression in the third line of (\ref{genvol}) with respect to $\lambda_a$.

Starting with the third line of (\ref{genvol}), we compute\footnote{It is perhaps worth noting that one 
cannot use Poincar\'e duality to replace the integral over $S_a$ in the expression on the 
second line of \eqref{deeveevee} as an integral over $X_{2r+1}$, simply because the integrand is not a closed form.}
\begin{align}\label{deeveevee}
\frac{\partial \mathcal{V}}{\partial v^i_a}
&=
{(2\pi)^{r+1}}\int_{\mathbb{R}^{r+1}}\delta((\vec{x},\v_a)-\lambda_a) \prod_{b\ne a} \theta((\vec{x},\v_b)-\lambda_b) 
\delta((\vec{x},\vec{b}))x_i\nn
&=(2\pi)\int_{S_a}x_i\eta\wedge\tfrac{1}{(r-1)!}\omega^{r-1}\, .
\end{align}
In the second equality we have rewritten the expression as an integral over $S_a$, as in (\ref{SaV}), where in doing so notice that $x_i$
may be interpreted as a Hamiltonian function for the transverse K\"ahler two-form $\omega$, as introduced in (\ref{Hamilton}).
Using the first line of \eqref{deeveevee} we also have
\begin{align}\label{beeten}
\sum_{a=1}^d\lambda^{a}\frac{\partial \mathcal{V}}{\partial v^i_a}&=
-\sum_{a=1}^d\lambda^a\frac{\partial}{\partial \lambda^a}\left[
{(2\pi)^{r+1}}\int_{\mathbb{R}^{r+1}}\prod_{b=1}^d \theta((\vec{x},\v_b)-\lambda_b) \delta((\vec{x},\vec{b}))x_i \right]\nn
&=-(r+1)\int_{\X_{2r+1}}x_i\eta\wedge\tfrac{1}{r!}\omega^{r}\, ,
\end{align}
where the second equality comes from the fact that the quantity in square brackets is homogeneous degree $r+1$ in the $\lambda_a$. 
More generally, we may similarly deduce
\begin{align}\label{moregen}
\sum_{a_1,\dots,a_s=1}^d\lambda^{a_1}\dots \lambda^{a_s}\frac{\partial^s \mathcal{V}}{\partial v^{i_1}_{a_1}\dots\partial v^{i_s}_{a_s}}&=
(-1)^s\left[\prod_{m=1}^s(r+m)\right]\int_{\X_{2r+1}}x_{i_1}\dots x_{i_s} \eta\wedge\tfrac{1}{r!}\omega^{r}\, .
\end{align}
From this, and similarly to (\ref{deeveevee}), we also deduce that 
\begin{align}
&  \sum_{a_1,\dots,a_s=1}^d\lambda^{a_1}\dots \lambda^{a_s}\frac{\partial^{s+1} \mathcal{V}}{\partial v^{i_1}_{a_1}\dots\partial v^{i_s}_{a_s}\partial v^i_{a}}\nonumber\\
 & \qquad = (2\pi) (-1)^s \left[\prod_{m=1}^s(r+m)\right]\int_{S_a}x_{i_1}\dots x_{i_s}x_i \eta\wedge\tfrac{1}{(r-1)!}\omega^{r-1}\, .
\end{align}

Next we would like to obtain expressions for derivatives of $\mathcal{V}$ with respect to the $R$-symmetry vector $\vec{b}$. These are obtained somewhat differently 
from the method above. First recall from section \ref{sec:setup} that
\begin{align}
\label{moments}
\partial_{\varphi_i}\lrcorner \eta \ = \ 2\sss_i~, \qquad 
\partial_{\varphi_i}\lrcorner \diff\eta \ = \ -2\diff \sss_i~, \qquad \partial_{\varphi_i}\lrcorner\omega = -\diff x_i~.
\end{align}
We may differentiate $\mathcal{V}$, defined by the first equality in \eqref{genvol}, with respect to the $R$-symmetry vector by 
taking the derivative inside the integral, and computing the corresponding first order variations of $\eta$ and $\omega$. Using the fact that the $R$-symmetry 
vector is
\begin{align}
\xi = \sum_{i=1}^{r+1} b_i\partial_{\varphi_i}\, ,
\end{align}
together with $\xi\lrcorner\eta=1$, $\xi\lrcorner\omega=0$ and \eqref{moments}, following section 4.2 of \cite{Gauntlett:2018dpc} one finds the first order variations
\begin{align}\label{variations}
\delta_{b_j}\eta &=-2w_j \eta+\nu^T_j\,,\nn
\delta_{b_j} \omega&=\eta \wedge \diff x_j-x_j \diff\eta+\diff\gamma_j^T\,,
\end{align}
where $\nu^T_j$ and $\gamma_j^T$ are basic forms for the foliation $\mathcal{F}_\xi$. Here in the variation of 
$\omega$ we hold the transverse K\"ahler class fixed, as in section 4.2 of \cite{Gauntlett:2018dpc}. 
In computing second derivatives of $\mathcal{V}$ we will also need the first order variations of $x_i$ and $w_i$. From \eqref{moments} and \eqref{variations} we may immediately deduce
\begin{align}
\delta_{b_j} w_i =-2 w_i w_j+\tfrac{1}{2}\partial_{\varphi_i}\lrcorner \nu_j^T\,.
\end{align}
Using the fact that the Lie derivatives of $x_j$ and $\gamma_j$ with respect to $\partial_{\varphi_i}$ vanish, we similarly find 
\begin{align}
\delta_{b_j} x_i =-2 w_i x_j+\partial_{\varphi_i}\lrcorner \gamma_j^T\,.
\end{align}

With these results to hand, using the first definition of $\mathcal{V}$ in \eqref{genvol}
it is straightforward to compute
\begin{align}
\frac{\partial \mathcal{V}}{\partial b_i}
& =
-\int_{\X_{2r+1}}\eta\wedge \left[
2w_i\frac{\omega^r}{r!}
+x_i\frac{\omega^{r-1}}{(r-1)!}\wedge \diff\eta\right]\,,\label{sdbibj}\\
\frac{\partial^2 \mathcal{V}}{\partial b_i\partial b_j} & =
\int_{\X_{2r+1}}\eta\wedge \left[
8w_iw_j\frac{\omega^r}{r!}+
8w_{(i} x_{j)} \frac{\omega^{r-1}}{(r-1)!}\wedge \diff\eta
+x_ix_j\frac{\omega^{r-2}}{(r-2)!}\wedge (\diff\eta)^2\right]\,.\nonumber
\end{align}
We may similarly take the derivative of the expression for 
$\sum_{a=1}^d\partial\mathcal{V}/\partial \lambda_a$ given by \eqref{lamderv} (with $s=1$) to obtain
\begin{align}\label{Vlambdab}
\sum_{a=1}^d\frac{\partial^2\mathcal{V}}{\partial\lambda_a\partial b_i}
 &=  b_1\int_{\X_{2r+1}}\eta\wedge \left[ 4w_i \frac{\omega^{r-1}}{(r-1)!}\wedge \diff\eta+
x_i \frac{\omega^{r-2}}{(r-2)!}\wedge (\diff\eta)^2  \right]\nn
& \qquad \qquad+\delta_{i1}\frac{1}{b_1}
\sum_{a=1}^d\frac{\partial\mathcal{V}}{\partial\lambda_a}\,,
\end{align}
where we have used $[\rho]=b_1[\diff\eta]\in H^2_B(\mathcal{F}_\xi)$ from \eqref{bcoh}. We shall also 
need
\begin{align}\label{mixeddersymm}
-\frac{\partial}{\partial b_j}\int_{\X_{2r+1}}\eta\wedge \tfrac{1}{r!}{\omega^r}x_i
& =\int_{\X_{2r+1}}\eta\wedge \left[ 4x_{(i}w_{j)}\tfrac{1}{r!}{\omega^r} + x_i x_j
\tfrac{1}{(r-1)!}{\omega^{r-1}}\wedge \diff\eta \right]\,,
\end{align}
and taking another derivative
\begin{align}
\frac{\partial^2}{\partial b_j\partial b_k}\int_{\X_{2r+1}}\eta\wedge \tfrac{1}{r!}{\omega^r} x_i  = & 
\int_{\X_{2r+1}}\eta\wedge \Big[ 24w_{(i}w_{j}x_{k)}\tfrac{1}{r!}{\omega^r}
+ 12w_{(i}x_j x_{k)}
\tfrac{1}{(r-1)!}{\omega^{r-1}}\wedge \diff\eta 
\nn
& \qquad \quad +x_ix_jx_k
\tfrac{1}{(r-2)!}{\omega^{r-2}}\wedge (\diff\eta)^2
\Big]\,.
\end{align}
Finally, we also have
\begin{align}
-\frac{\partial}{\partial b_{k}}\int_{\X_{2r+1}}\eta\wedge 
\tfrac{1}{r!}{\omega^r} x_{i}x_{j}
& =\int_{Y_{2r+1}}\eta\wedge \Big[ 6w_{(i}x_{j}x_{k)}
\tfrac{1}{r!}{\omega^r} \nn & \qquad \qquad+ x_{i} x_{j}x_{k}
\tfrac{1}{(r-1)!}{\omega^{r-1}}\wedge \diff\eta \Big]\,.
\end{align}

Turning now to  integrals over the toric codimension two submanifolds $S_a\subset \X_{2r+1}$,
taking the derivative of the expression given in \eqref{SaVnew} we find
\begin{align}\label{SaVnewagain}
\frac{(-1)^s}{2\pi}\sum_{b_1,\dots,b_s=1}^d \frac{\partial^{s+2} \mathcal{V}}{\partial b_i \partial \lambda_a\partial \lambda_{b_1}\dots\partial\lambda_{b_s}}& =\int_{S_a} \eta\wedge \rho^s\wedge \Big[2(1+s)w_i\tfrac{1}{(r-s-1)!}\omega^{r-s-1} \nn 
& \qquad \qquad +
x_i \diff\eta\wedge \tfrac{1}{(r-s-2)!}\omega^{r-s-2}\Big]
\,.
\end{align}
In particular for $s=0$ we have
\begin{align}\label{SaVnewag}
\frac{1}{2\pi} \frac{\partial^{2} \mathcal{V}}{\partial b_i \partial \lambda_a}=
\int_{S_a} \eta\wedge \Big[2w_i\tfrac{1}{(r-1)!}\omega^{r-1}+
x_i \diff\eta\wedge \tfrac{1}{(r-2)!}\omega^{r-2}\Big]\,.
\end{align}
We similarly have
\begin{align}\label{bill}
-\frac{\partial}{\partial b_j}\int_{S_a}\eta\wedge \tfrac{1}{(r-1)!}\omega^{r-1} x_i
& =\int_{S_a}\eta\wedge  \Big[  4w_{(i}x_{j)}\tfrac{1}{(r-1)!}\omega^{r-1} \nn
& \qquad \qquad +x_i x_j \diff\eta\wedge \tfrac{1}{(r-2)!}\omega^{r-2} \Big]\,,
\end{align}
as well as
\begin{align}
-\frac{\partial}{\partial b_{k}}\int_{S_a}\eta\wedge \tfrac{1}{(r-1)!}\omega^{r-1} x_{i}x_{j}
& =\int_{S_a}\eta\wedge  \Big[ 6w_{(i}x_{j}x_{k)}\tfrac{1}{(r-1)!}\omega^{r-1} \nn
& \qquad \qquad +x_{i} x_{j}x_{k}\diff\eta\wedge \tfrac{1}{(r-2)!}\omega^{r-2} \Big]\,.
\end{align}

\subsection{Formulae for $\X_{2r+1}\hookrightarrow Y_{2r+5}\rightarrow B_4$}\label{app:beefour}

In this subsection we explain how to derive the formulae presented in section \ref{beefour}, starting from 
the general expressions for the constraint (\ref{constraint}), flux quantization conditions (\ref{quantize}) and supersymmetric 
action (\ref{susyact}). 

Starting with the supersymmetric action \eqref{susyact}, 
making the substitutions in \eqref{subst} 
and recalling \eqref{bcoh} immediately gives
\begin{align}
\Ssusy & = \int_{Y_{2n+1}}\eta\wedge b_1\left(\diff\eta+2w_iF_i\right)\wedge \tfrac{1}{(n-1)!}\left(\omega+x_jF_j+J_{B_{4}}\right)^{n-1}\,\nn
& = \int_{Y_{2n+1}}\eta\wedge b_1\diff\eta\wedge \tfrac{1}{(n-3)!}\omega^{n-3}\wedge \tfrac{1}{2}J_{B_{4}}\wedge J_{B_{4}}\nn
& \quad + b_1\int_{Y_{2n+1}}\eta\wedge\left[2w_i\tfrac{1}{(n-2)!}\omega^{n-2} + x_i\diff\eta\wedge \tfrac{1}{(n-3)!}\omega^{n-3}\right]\wedge F_i\wedge J_{B_{4}}\nn
& \quad + b_1\int_{Y_{2n+1}}\eta\wedge \left[4w_{(i}x_{j)}\tfrac{1}{(n-2)!}\omega^{n-2} + x_i x_j \diff\eta\wedge\tfrac{1}{(n-3)!}\omega^{n-3}\right]\wedge \tfrac{1}{2}F_i\wedge F_j\, .
\end{align}
Here in the second equality we have simply collected terms together and written them as a $(2r+1)$-form  on the fibre 
$\X_{2r+1}$ wedged with a four-form on the base $B_4$, where $n=r+2$. This then leads to the elegant expression 
\begin{align}\label{SsusyB4old}
\Ssusy&=
 -\sum_{a=1}^d\frac{\partial\mathcal{V}}{\partial\lambda_a}\vol(B_4) - b_1\sum_{i=1}^{r+1}\frac{\partial\mathcal{V}}{\partial b_i}\int_{B_4}F_i\wedge J_{B_{4}}  \nn 
& \qquad + \frac{b_1}{r+1}\sum_{a=1}^d\sum_{i,j=1}^{r+1}\lambda_a\frac{\partial^2\mathcal{V}}{\partial b_j\partial v_a^i}\int_{B_4}\tfrac{1}{2}F_i\wedge F_j\, ,
\end{align}
where $\vol(B_4)=\int_{B_4}\tfrac{1}{2}J_{B_{4}}\wedge J_{B_{4}}$. 
Here we have used equation \eqref{lamderv} (with $s=1$) for the first term, equation \eqref{sdbibj} for the second term, 
and equation \eqref{mixeddersymm} for the third term. The final form of the third term presented in \eqref{SsusyB4old} 
then further uses the expression \eqref{beeten} for the left hand side of \eqref{mixeddersymm}.

Next we turn to the constraint equation (\ref{constraint}).  Using \eqref{subst} 
and \eqref{bcoh}
this similarly expands as
\begin{align}
0 &= \int_{Y_{2n+1}}\eta\wedge b_1\left(\diff\eta+2w_iF_i\right)\wedge b_1\left(\diff\eta + 2w_jF_j\right)\wedge \tfrac{1}{(n-2)!}\left(\omega+x_kF_k+J_{B_{4}}\right)^{n-2}\nn
& = \int_{Y_{2n+1}}\eta\wedge \rho^2\wedge\tfrac{1}{(n-4)!}\omega^{n-4}\wedge \tfrac{1}{2}J_{B_{4}}\wedge J_{B_{4}} \nn
& \quad + b_1^2\int_{Y_{2n+1}} \eta\wedge\Big[x_i(\diff\eta)^2\wedge\tfrac{1}{(n-4)!}\omega^{n-4} + 4w_i\diff\eta\wedge \tfrac{1}{(n-3)!}\omega^{n-3}\Big]\wedge F_i\wedge J_{B_{4}}\nn 
& \quad + b_1^2\int_{Y_{2n+1}}\eta\wedge\Big[8w_iw_j\tfrac{1}{(n-2)!}\omega^{n-2}+8w_{(i}x_{j)}\diff\eta\wedge \tfrac{1}{(n-3)!}\omega^{n-3}\nn
& \qquad\qquad\qquad\qquad\qquad \qquad\qquad +x_ix_j (\diff\eta)^2\wedge \tfrac{1}{(n-4)!}\omega^{n-4}\Big]\wedge\tfrac{1}{2}F_i\wedge F_j\, ,
\end{align}
where again $n=r+2$. 
We may then use equation \eqref{lamderv} (with $s=2$) for the first term, equation \eqref{Vlambdab} for the second term, 
and equation \eqref{sdbibj}  for the third term. This immediately gives
\begin{align}
&\sum_{a,b=1}^d  \frac{\partial^2\mathcal{V}}{\partial\lambda_a\partial\lambda_b} \vol(B_4)
+b_1\sum_{i=1}^{r+1}\sum_{a=1}^d \frac{\partial^2\mathcal{V}}{\partial\lambda_a\partial b_i}\int_{B_4}F_i\wedge J_{B_{4}}
-\sum_{a=1}^d \frac{\partial\mathcal{V}}{\partial\lambda_a}\int_{B_4}F_1\wedge J_{B_4}\nn
&\qquad+{b_1^2}\sum_{i,j=1}^{r+1}\frac{\partial^2 \mathcal{V}}{\partial b_i\partial b_j}\int_{B_4}\tfrac{1}{2}F_i\wedge F_j = 0\, ,
\end{align}
which is the constraint equation \eqref{constraintB4} presented in the main text. 

For the flux quantization condition there are two types of cycle. The first type has the fibred form 
$\X_{2r+1}\hookrightarrow \Sigma_\alpha \rightarrow C_\alpha^{(2)}$, where $C_\alpha^{(2)}\subset B_4$ is a two-cycle. 
In this case the flux quantization condition \eqref{quantize} reads
\begin{align}
\nu_n N_\alpha & = \int_{\Sigma_\alpha}\eta\wedge b_1(\diff\eta + 2w_iF_i)\wedge \tfrac{1}{(n-2)!}\left(\omega+ x_jF_j+J_{B_4}\right)^{n-2}\nn
& = \int_{\Sigma_\alpha}\eta\wedge b_1\diff\eta \wedge\tfrac{1}{(n-3)!}\omega^{n-3}\wedge J_{B_4} 
\nn
&\quad+ b_1\int_{\Sigma_\alpha}\eta\wedge \Big[2w_i\tfrac{1}{(n-2)!}\omega^{n-2} 
x_i\diff\eta\wedge \tfrac{1}{(n-3)!}\omega^{n-3}\Big]\wedge F_i\, .
\end{align}
Using equation \eqref{lamderv} (with $s=1$) for the first term, and equation \eqref{sdbibj} for the second term then leads to
\begin{align}\label{NalphaB4app}
\nu_n N_\alpha &= -\sum_{a=1}^d  \frac{\partial  \mathcal{V}}{\partial \lambda_a } \int_{C^{(2)}_\alpha}J_{B_4}
-b_1\sum_{i=1}^{r+1}\frac{\partial  \mathcal{V}}{\partial b_i }\int_{C^{(2)}_\alpha}F_i
\,,
\end{align}
which is equation \eqref{NalphaB4} in the main text. 
The second set of cycles have the fibred form $S_a\hookrightarrow \Sigma_a\rightarrow B_4$, 
where $S_a\subset \X_{2r+1}$ is a toric codimension two submanifold in the fibre. 
In this case the flux quantization condition \eqref{quantize} reads
\begin{align}
\nu_n M_a & = \int_{\Sigma_a}\eta\wedge b_1(\diff\eta + 2w_iF_i)\wedge \tfrac{1}{(n-2)!}\left(\omega + x_jF_j + J_{B_4}\right)^{n-2}\nn
& = b_1\int_{\Sigma_a}\eta\wedge \diff\eta \wedge \tfrac{1}{(n-4)!}\omega^{n-4}\wedge \tfrac{1}{2}J_{B_4}\wedge J_{B_4} \nn 
& \quad + b_1\int_{\Sigma_a}\eta\wedge \Big[x_i\diff\eta\wedge \tfrac{1}{(n-4)!}\omega^{n-4} + 2w_i\tfrac{1}{(n-3)!}\omega^{n-3}\Big]\wedge F_i\wedge J_{B_4}\nn
& \quad + b_1\int_{\Sigma_a}\eta\wedge\Big[x_ix_j\diff\eta \wedge \tfrac{1}{(n-4)!}\omega^{n-4} + 4w_{(i}x_{j)}\tfrac{1}{(n-3)!}\omega^{n-3}\Big]\wedge \tfrac{1}{2}F_i\wedge F_j\, ,
\end{align}
where the flux quantum number is denoted $M_a\in\Z$. Using equation \eqref{SaVnew} (with $s=1$) for the first term, equation \eqref{SaVnewag} for the second term, and 
combining equations \eqref{deeveevee} and \eqref{bill}  
for the third term 
then leads to
\begin{align}\label{MaB4app}
\nu_{n} M_a & = \frac{1}{2\pi}\sum_{b=1}^d \frac{\partial^2\mathcal{V}}{\partial\lambda_a\partial\lambda_b}\vol(B_4)
+\frac{b_1}{2\pi}\sum_{i=1}^{r+1}\frac{\partial^2\mathcal{V}}{\partial\lambda_a\partial b_i} \int_{B_4}F_i\wedge J_{B_4}\nn
&\qquad-\frac{b_1}{2\pi}\sum_{i,j=1}^{r+1}
\frac{\partial^2 \mathcal{V}}{\partial {b_{j}}\partial v^{i}_a}
\int_{B_4}\tfrac{1}{2}F_{i}\wedge F_{j}\,,
\end{align}
which is equation \eqref{MaB4} in the main text. 

Finally, the form of the supersymmetric action presented in \eqref{SsusyB4} may be obtained from \eqref{SsusyB4old}
by first summing \eqref{MaB4app} over $a=1,\ldots,d$, and then using \eqref{NalphaB4app} together with the fact
that the master volume $\mathcal{V}$ is homogeneous degree $r$ in the $\lambda_a$. 
The formulae presented in sections \ref{beetwo} and \ref{beesix} are obtained in an entirely analogous manner, in particular using 
the equations we have so far not used in Appendix \ref{app:der}.

\section{Explicit solutions with K\"ahler-Einstein factors}\label{app:examples}

In this appendix we present some explicit solutions of type IIB and $D=11$
supergravity where the K\"ahler base $B_{2\Bp}$
is K\"ahler-Einstein with positive curvature. The results from this section 
are compared with some of the results obtained using our general formalism in section \ref{sec:examples}.

\subsection{Type IIB: $\mathscr{Y}^{\pJ,\qJ}(KE_4^+)$}\label{iibexpl}

We first recall the class of explicit $AdS_3\times Y_7$ solutions
of type IIB supergravity of the form  \eqref{ansatz} that were constructed in \cite{Gauntlett:2006af}.
The solutions, which we label $Y_7=\mathscr{Y}^{\pJ,\qJ}(KE_4^+)$,
are constructed using an arbitrary K\"ahler-Einstein manifold with
positive curvature, $KE_4^+$, and are specified by two positive, relatively prime\footnote{This ensures $Y_7$ is simply connected.} integers,
$\pJ,\qJ>0$. As
explained in detail in \cite{Gauntlett:2006ns}, $\mathscr{Y}^{\pJ,\qJ}(KE_4^+)$ can be constructed
as a circle fibration over a regular six-dimensional manifold, which itself is obtained
by constructing an $S^2$ bundle over $KE_4^+$.  Equivalently, 
$\mathscr{Y}^{\pJ,\qJ}(KE_4^+)$ can also be viewed as the total space of a Lens space $L(\qJ,1)=S^3/\Z_\qJ$
fibred over $KE_4^+$.

The analysis of the regularity of the solutions, flux quantization, and calculation of the central charge was carried out in detail in \cite{Gauntlett:2006af}. In the notation of
\cite{Gauntlett:2006af} the flux integrals are given by
\begin{align}\label{finalNIIB}
N(D_0)&=-\frac{ M}{hm^2}(\pJ+m\qJ) n\,,  \nn
N(\tilde D_0)&=-\frac{M}{hm^2}\pJ n \,, \nn
N(D_a)&=\frac{\qJ}{h} n_a n \,,
\end{align}
while the central
charge is given by
\begin{align}\label{iibexplccfin}
c=\frac{9\pJ\qJ^3(\pJ+m\qJ)}{3\pJ^2+3m\pJ\qJ+m^2 \qJ^2}\frac{M}{m^2h^2}n^2\,,
\end{align}
where $n$ is an arbitrary integer. In addition, the integers $m$ and $M$ depend on the specific choice of $KE_4^+$.
If $\mathcal{K}$ is the canonical line bundle of the $KE_4^+$ then the 
Fano index $m$ is the largest
positive integer $m$ for which there is a line bundle $\mathcal{N}$ with $\mathcal{K}=\mathcal{N}^m$.
If $\rho_{KE}$ is the Ricci-form of the $KE_4^+$ then $M=\int_{KE_4^+}(\frac{1}{2\pi}\rho_{KE})^2=\int_{KE_4^+}c_1^2$.
For
$S^2\times S^2$ we then have $m=2$, $M=8$; for $\C P^2$ we have $m = 3$, $M = 9$; 
and for the del Pezzo surfaces  $dP_k$, we have $m = 1$, $M = 9 - k$, where $k=3,\ldots,8$. Finally, $h = \mathrm{hcf}(M/m^2, \qJ)$.

\subsection{$D=11$: $\mathscr{Y}^{\pJ,\qJ}(KE_6^+)$}\label{appdd11}

We next discuss a class of explicit $AdS_2\times Y_9$ solutions
of $D=11$ supergravity of the form \eqref{ansatzd11}, with $Y_9=\mathscr{Y}^{\pJ,\qJ}(KE_6^+)$.
These solutions were first discussed in section~3.2 of \cite{Gauntlett:2006ns}, including determining the
conditions required to obtain regular solutions, which are labelled by two positive, relatively prime integers, $\pJ,\qJ>0$.
Here we will
carry out flux quantization and obtain an explicit expression for the entropy. This will
enable us to compare, successfully, with the general formalism of this paper in section \ref{d11exsolcase}.

The metric takes the form
\begin{equation}
\label{q-met}
\begin{aligned}
   \diff s^2(Y_9) = \frac{y^3-3y+2a}{y^3}Dz^2
     + \frac{4\diff y^2}{q(y)}+\frac{q(y)(D\psi)^2}{y^3(y^3-3y+2a)}
      + \frac{16}{y^2}\diff s^2(KE_6^+)\,,
\end{aligned}
\end{equation}
where $\diff s^2(KE_6^+)$ is an arbitrary six-dimensional K\"ahler-Einstein metric, normalized
so that $\rho_{KE}=8J_{KE}$. Moreover, we have introduced the functions
\begin{equation}
\begin{aligned}
   q(y) = y^4 - 4 y^2 + 4 a y - a^2 \,,\qquad
   g(y) = \frac{a-y}{y^3 - 3 y + 2a} \,,
\end{aligned}
\end{equation}
and covariant derivatives
\begin{align}
Dz\equiv \diff z-g(y)D\psi\, , \qquad D\psi\equiv \diff\psi+4B\, , 
\end{align}
where $4B$ is the natural connection on the canonical line bundle of $KE_6^+$ (i.e. $\diff B=2J_{KE}$).  
The constant $a$, explicitly given in terms of $\pJ,\qJ$ below, lies in the range $0<a<1$.
The quartic $q(y)$ then has four distinct roots, $y_1<y_2<y_3<y_4$, and we choose 
the range of $y$ to be $y_2\le y\le y_3$ where
\begin{align}
y_2=-1+\sqrt{1+a}\,,\qquad
y_3=1-\sqrt{1-a}\,.\
\end{align}
Potential conical singularities at $y=y_2,y_3$ are avoided by taking $\psi$ to be a periodic coordinate with period $2\pi$. Finally, as explained in \cite{Gauntlett:2006ns}, the coordinate 
$z$ is periodic with period $2\pi l$, with
\be
g(y_3)-g(y_2)=l\qJ,\qquad g(y_2)=l\pJ/m\,,
\ee
where $\pJ,\qJ>0$ are relatively prime\footnote{This condition ensures that
$Y_9$ is simply connected.} integers
and the integer $m$ is the Fano index of $KE_6^+$.
These conditions are satisfied  provided that
\begin{align}
a&=\frac{m\qJ(2\pJ+m\qJ)}{2\pJ^2+2m\pJ \qJ+m^2\qJ^2}\,,\nn
l&=\frac{m[\pJ^2+m\pJ \qJ+(m^2/2)\qJ^2]^{1/2}}{\pJ(\pJ+m\qJ)}\,.
\end{align}

We now turn to flux quantization. We begin by noting that the 
two-form $F$, entering into the expression for the four-form $G$ in \eqref{ansatzd11},
is given by
\be
F=\frac{2^{3/2}}{3^{3/2}}\left(3y^2\diff y\wedge \diff z-8aJ_{KE}\right)\,.
\ee
After taking the $D=11$ Hodge dual of $G$ we obtain 
the seven-form given by 
\begin{align}
L^{-6}*_{11}G & = \frac{2^{14}}{3^2}\frac{q(y)}{y(y^3-3y+2a)}\frac{1}{3!}J^3_{KE}\wedge D\psi
+\frac{2^{14}}{3^2}\frac{y-a}{y}\frac{1}{3!}J^3_{KE}\wedge Dz\nn
&\quad +\frac{a}{y^2} \frac{2^{11}}{3^3}\frac{1}{2!}J^2_{KE}\wedge \diff y\wedge D\psi\wedge Dz\,.
\end{align}
Flux quantization requires that
\begin{align}
\frac{1}{(2\pi \ell_p)^6}\int_{\Sigma_A} *_{11}G &= N_A \in \mathbb{Z}~,\label{quantization11}
\end{align}
over all seven-cycles $\Sigma_A\subset Y_9$,
where $A=1,\ldots, \mathrm{rank}\, H_7(Y_9,\Z)$ runs over an integral basis for the free part of $H_7(Y_9,\Z)$, and 
$\ell_p$ denotes the eleven-dimensional Planck length. 

To proceed further we need a basis for the free part of $H_7 (Y_9, \mathbb{Z})$. 
Recall that $Y_9$ is the total space of $U(1)$ fibration, with fibre  coordinate $z$, over an eight-dimensional manifold, $B_8$, the latter being the total space of an
$S^2$ bundle over $KE_6^+$. 
A basis for the free part of $H_6 (B_8 , \mathbb{Z})$ is given by a section of the $S^2$ bundle over $KE_6^+$, 
say at $y_2$ or $y_3$, together with the total spaces of the $S^2$ fibrations over each basis four-cycle $\Sigma_a \in H_4(KE_6^+,\mathbb{Z})$. 
It will be useful in a moment to write the 
Poincar\'e dual of the first Chern class of the $m$th root of the canonical line bundle of $KE_6^+$, denoted  $c_1(\mathcal{N})$,
as $[c_1(\mathcal{N})] =s_a \Sigma_a$, where $s_a$ are a set of co-prime integers.

Now,
since the $U(1)$ bundle over $B_8$ is non-trivial, all non-trivial seven-cycles come from the total space of the
$U(1)$ fibration over a six-cycle in $B_8$. Let us label these as follows: $D_0$ denotes the seven-cycle 
arising from the section $y = y_2$, $\tilde D_0$ is the cycle corresponding to $y = y_3$, and $D_a$ the cycle arising from $\Sigma_a$. Note that these cycles are not independent. From the $S^2$ fibration structure of $B_8$ we have
\begin{align}
D_0 & = \tilde D_0 - ms_a D_a\, ,
\label{homoone}
\end{align}
while the $U(1)$ fibration is such that 
\begin{align}
0 & = \qJ\tilde D_0 + \pJ s_aD_a \, .
\label{homotwo}
\end{align}
The flux integrals are then explicitly given by 
\begin{align}
N(D_0)&=-\frac{2^2}{ 3^3 \pi^2}\left(\frac{L}{\ell_p}\right)^6 \frac{mM}{\pJ} \,,  \nn
N(\tilde D_0)&=-\frac{2^2}{ 3^3 \pi^2}\left(\frac{L}{\ell_p}\right)^6 \frac{mM}{\pJ+m\qJ}\,, \nn
N(D_a)&=\frac{2^2}{ 3^3 \pi^2}\left(\frac{L}{\ell_p}\right)^6 \frac{m^4\qJ }{\pJ(\pJ+m\qJ)} n_a \,,
\end{align}
where
\begin{align}
M & \equiv \int_{KE_6^+}\left(\frac{\rho_{KE}}{2\pi}\right)^3 = \int_{KE_6^+}c_1^3\,,
\end{align}
and we have also used 
\begin{align}
\int_{\Sigma_a}\left(\frac{\rho_{KE}}{2\pi}\right)^2=m^2 n_a\,,
\end{align}
for some co-prime integers $n_a \equiv \int_{\Sigma_a} c_1(\mathcal{N})^2$,
which follows from
\begin{align}
 \frac{\rho_{KE}}{2\pi} & = c_1 = m c_1(\mathcal{N})\, ,
\end{align}
with $m^3 s_a n_a=M$.
Thus we should choose
\begin{align}
\left(\frac{L}{\ell_p}\right)^6=\frac{3^3\pi^2}{2^2}\frac{\pJ(\pJ+m \qJ) n}{h m^4}\, ,
\end{align}
where $h=\mathrm{hcf}(M/m^3,\qJ)$ and $n$ is an integer, so that
\begin{align}\label{finalNsd11}
N(D_0)&=-\frac{ M}{hm^3}(\pJ+m\qJ) n\,,  \nn
N(\tilde D_0)&=-\frac{M}{hm^3}\pJ n \,, \nn
N(D_a)&=\frac{\qJ}{h} n_a n \, , 
\end{align}
are all integers. Notice these are consistent with the homology relations (\ref{homoone}) and (\ref{homotwo}).

With these ingredients to hand, 
and using  \cite{Gauntlett:2006ns}
\begin{align}
\ex^{B} & = \frac{R}{2} = \left(\frac{3}{2}\right)^{3/2} \frac{1}{y^3}\, , 
\end{align}
where $R$ is the Ricci scalar of the eight-dimensional transverse K\"ahler metric,
we may now compute the 
``entropy" given by \eqref{cS2}
(see equation (2.20) of \cite{Couzens:2018wnk})
\begin{align}\label{finalentd11}
\mathscr{S}&=\frac{1}{2^6\pi^7}\left(\frac{L}{\ell_p}\right)^9\int_{Y_9} \ex^{-3B}\vol_9\,,\nn
&=\frac{2^{1/2}}{3^{1/2}}\frac{2^6}{3^5\pi^2}\left(\frac{L}{\ell_p}\right)^9(y_3-y_2)l M\,,\nn
&=
\frac{2^{7/2} \pi  M \sqrt{\pJ (m \qJ+\pJ)} \left[
 \sqrt{(m \qJ+2 \pJ)^2+m^2 \qJ^2}-(m \qJ+2 \pJ) \right]}
{3 m^5 h^{3/2}}n^{3/2}\,.
\end{align}

\section{Sasakian volume function}\label{app:Sas}

As originally pointed out in \cite{Gauntlett:2007ts}, GK geometry shares many similarities with Sasakian geometry. In this appendix we point out that the formalism developed 
in \cite{Gauntlett:2019roi, Gauntlett:2018dpc} and the present paper allows one to efficiently compute the Sasakian volume function of \cite{Martelli:2005tp, Martelli:2006yb} 
in many interesting cases. 

Recall that a Sasakian manifold $(Y_{2n+1},\diff s^2_{2n+1})$ of real dimension $2n+1$ with $n\ge 1$, 
may be defined as a Riemannian manifold whose metric cone \eqref{metriccone} is K\"ahler. 
Precisely as for GK geometry in section \ref{sec:GKsummary}, there is unit norm Killing vector $\xi$ on $Y_{2n+1}$ with dual one-form $\eta$, so that the Sasakian metric may be written as
\begin{align}
\diff s^2_{2n+1} = \eta^2 + \diff s^2_{2n}\, ,
\end{align}
where $\diff s^2_{2n}$ is a K\"ahler metric transverse to the foliation $\mathcal{F}_\xi$ generated by $\xi$, c.f. \eqref{GKmetric} for GK geometry. In Sasakian geometry 
$\eta$ is a contact one-form, satisfying $\diff\eta = 2J$, where $J$ is the transverse K\"ahler form. 

If one is interested in Sasaki-Einstein 
metrics, where the metric cone \eqref{metriccone} is Ricci-flat K\"ahler, then there is necessarily a nowhere zero holomorphic $(n+1,0)$-form 
$\Psi$ on the cone satisfying
\begin{align}\label{Reebcharge}
\mathcal{L}_\xi \Psi = \ii (n+1) \Psi\, .
\end{align}
Suppose that \eqref{Reebcharge} holds on the K\"ahler cone. 
As in section \ref{sec:toric} we may write the Reeb vector as
\begin{align}
\xi = \sum_{i=1}^{r+1} b_i \partial_{\varphi_i}\, 
\end{align}
where by definition $\Psi$ has unit charge under $\partial_{\varphi_1}$, and is uncharged under $\partial_{\varphi_i}$, $i=2,3,\ldots, r+1\geq 1$. 
The condition \eqref{Reebcharge} then implies that 
\begin{align}\label{etaScoh}
[\diff\eta] = \frac{1}{b_1}[\rho]\in H^2_B(\mathcal{F}_\xi)\, ,
\end{align}
precisely as in \eqref{bcoh}, except that for Sasakian geometry we should then set $b_1=n+1$  so that \eqref{Reebcharge} holds. 

If we now define the Sasakian volume
\begin{align}
\mathrm{Vol}(Y_{2n+1}) \equiv  \int_{Y_{2n+1}} \eta\wedge \frac{J^n}{n!}\, ,
\end{align}
where $\diff\eta=2J$ and \eqref{etaScoh} holds, then the above comments imply that
\begin{align}\label{volSsusy}
\mathrm{Vol}(Y_{2+1}) = \frac{1}{2nb_1}\Ssusy\left(\xi, [J]= \tfrac{1}{2b_1}[\rho]\right)\, ,
\end{align}
where $\Ssusy$ is the supersymmetric action \eqref{susyact}, and one should set $b_1=n+1$. 

With the general formula \eqref{volSsusy} in hand, we may now compute the Sasakian volume function in the case that 
$Y_{2n+1}$ is the total space of a toric $\X_{2r+1}$ fibration over a K\"ahler base $B_{2\Bp}$, where $n=r+\Bp$, and 
with the Reeb vector $\xi$ tangent 
to the fibres. We present formulae for $\Bp=1$ and $\Bp=2$ below, together with some illustrative examples.
We also note that the formulae are valid for Sasakian geometry on $Y_{2n+1}$ with $n\ge 1$ (even though the analysis in the bulk of the paper was for GK geometry with $n\ge 3$).

\subsubsection*{Base $B_2$}

Taking $\Bp=1$, the base $B_2$ is a Riemann surface. However, in Sasakian geometry 
where \eqref{Reebcharge} holds the 
K\"ahler class is a positive multiple of the anti-canonical class, which implies that $B_2=S^2$ necessarily has genus zero. 
In this case the formalism 
in section~\ref{beetwo} (generalizing \cite{Gauntlett:2019roi, Gauntlett:2018dpc}) gives the general formula
\begin{align}\label{Volbeetwo}
\mathrm{Vol}(Y_{2n+1}) = - \left. \frac{1}{2nb_1}\left[\frac{1}{b_1}\sum_{a=1}^d \frac{\partial\mathcal{V}}{\partial\lambda_a} + 2\pi b_1\sum_{i=1}^{r+1} n_i \frac{\partial\mathcal{V}}{\partial b_i}\right]\right|_{\lambda_a = -\frac{1}{2b_1},\, b_1=n+1}\, ,
\end{align}
where $n=r+1$.
Here as usual $\mathcal{V}$ denotes the master volume of the  $\X_{2r+1}$ fibres, and 
we have defined
\begin{align}
n_i \equiv \frac{1}{2\pi}\int_{S^2} F_i\, ,
\end{align}
which describes the twisting of the  fibres over the base $B_2=S^2$. 
As for GK geometry, the existence of a holomorphic $(n+1,0)$-form $\Psi$ on the metric cone over $Y_{2n+1}$ implies that $n_1=2$, which is the anti-canonical 
class of $S^2$ in $\Z\cong H^2(S^2,\Z)$. On the right hand side of \eqref{Volbeetwo} one should set all $\lambda_a=-\frac{1}{2b_1}$, $a=1,\ldots,d$, 
after taking derivatives with respect to the $\lambda_a$, and also $b_1=n+1$, after taking derivatives with respect to the $b_i$. 
The former condition ensures that the K\"ahler class of the fibres satisfies $[\omega]=\frac{1}{2b_1}[\rho]$, as in \eqref{lambdaSas}. 

To illustrate \eqref{Volbeetwo}, let us consider the case of three-dimensional fibres, with $r=1$. In this case $Y_5$ is the total space 
of a Lens space $X_3\cong S^3/\Z_p$ fibration over $S^2$, where the master volume $\mathcal{V}$ of the fibres 
is given by \eqref{VolX3}. Using \eqref{Volbeetwo} we easily compute the Sasakian volume
\begin{align}\label{Y5vol}
\mathrm{Vol}(Y_5) = \frac{p [2 b_2 (p-n_2)+3  p n_2]}{b_2^2 (b_2-3 p)^2}\pi ^3 \, .
\end{align}
Setting the twisting variable $n_2=p+q$, with\footnote{The inequality comes from requiring the metric cone over $Y_5$ to be an 
affine variety. This is perhaps easiest to see using toric geometry, since $Y_5$ is here toric. In terms of the fibration picture described in the present paper, 
recall that $X_3\cong S^3/\Z_p$ is the link of the $A_{p-1}$ singularity $C(X_3)=\C^2/\Z_p$. The corresponding $A_{p-1}$ fibration 
over $S^2$ is then a partial resolution of the affine cone $C(Y_5)$. The two divisors $z_1=0$ and $z_2=0$ in the fibre $\C^2/\Z_p$
give rise to $\C/\Z_p$ fibrations over $S^2$, with Chern numbers $-n_2=-(p+q)$ and $-2p+n_2=-(p-q)$. Both should be negative, in order 
that the total spaces are holomorphically convex. For more general examples there will be similar convexity conditions on the 
twisting parameters that need to be imposed.} $p>q>0$, extremizing \eqref{Y5vol}
over $b_2$ one finds that the unique critical point inside the Reeb cone is
\begin{align}
b_2 = \frac{p\left(2 p+3 q-\sqrt{4 p^2-3 q^2}\right)}{2 q}\, .
\end{align}
The on-shell volume is then
\begin{align}
\mathrm{Vol}(Y_5) = \frac{q^2 \left(2 p + \sqrt{4 p^2-3 q^2}\right)}{3 p^2 \left(-2 p^2+3 q^2 + p \sqrt{4 p^2-3 q^2}\right)}\pi^3\, ,
\end{align}
which agrees with the volume of the $Y^{p,q}$ Sasaki-Einstein manifolds \cite{Gauntlett:2004yd}, as expected.
 
\subsubsection*{Base $B_4$}

Taking $\Bp=2$, the base $B_4$ is now a K\"ahler surface. Again, this should be Fano, having positive 
anti-canonical class. In this case we find
\begin{align}\label{Volbeefour}
\mathrm{Vol}(Y_{2n+1}) & = - \frac{1}{2nb_1}\Bigg[\frac{\pi^2}{2b_1^2}\sum_{a=1}^d\frac{\partial\mathcal{V}}{\partial\lambda_a}\int_{B_4}c_1^2 + \pi \sum_{i=1}^{r+1}\frac{\partial\mathcal{V}}{\partial b_i}\int_{B_4}F_i\wedge c_1 \nn
& \left. - \frac{b_1}{r+1}\sum_{a=1}^d\sum_{i,j=1}^{r+1}\lambda_a\frac{\partial^2\mathcal{V}}{\partial b_j\partial v_a^i}\int_{B_4}\tfrac{1}{2}F_i\wedge F_j\Bigg]\right|_{\lambda_a = -\frac{1}{2b_1},\, b_1=n+1}\, ,
\end{align}
where now $n=r+2$. As in GK geometry we have $[F_1] = 2\pi c_1\in H^2(B_4,\R)$, where $c_1=c_1(B_4)$ is the anti-canonical class of the base $B_4$.  

Again, we illustrate \eqref{Volbeefour} by taking three-dimensional fibres, with $r=1$, and choose the base to be $B_4=\mathbb{C}P^2$. 
In this case there is a single generator of the cohomology $H=1\in\Z\cong H^2(\mathbb{C}P^2,\Z)$, given by the hyperplane class, with $\int_{\mathbb{C}P^2}H^2=1$. 
Writing $[F_2]=2\pi \kappa H$ with $\kappa\in\Z$, and using $c_1=3H$, we compute
\begin{align}
\mathrm{Vol}(Y_7) = \frac{p \left[\kappa^2 \left(16 p^2-12 b_2 p+3 b_2^2\right)+3 b_2 \kappa p (4 p-3 b_2)+9 p^2b_2^2 \right]}{3 b_2^3 (4p-b_2)^3}\pi ^4 \, .
\end{align}
Extremizing over $b_2$, we find that the on-shell volume agrees with the $Y^{p,\kappa}(\mathbb{C}P^2)$ Sasaki-Einstein manifolds of \cite{Gauntlett:2004hh}\footnote{See also \cite{Martelli:2008rt}, 
where the $Y^{p,\kappa}(\mathbb{C}P^2)$ notation was more precisely defined. We note that there is a typographical error in the volume in equation (2.23) of that paper, where the right hand side 
should be multiplied by $\tfrac{1}{4}$ to obtain the Sasaki-Einstein volume with standard normalization we are using in this appendix.}, 
as expected.


\providecommand{\href}[2]{#2}\begingroup\raggedright\endgroup

\end{document}